\documentclass[twocolumn,showpacs,preprintnumbers,amsmath,amssymb]{revtex4}

\usepackage{epsfig}
\usepackage{subfigure}
\usepackage{bm}
\usepackage{graphicx}
\usepackage{dcolumn}
\usepackage{amsmath}
\graphicspath{{pics/}}    

\usepackage[normalem]{ulem} 
\usepackage[dvips]{color} 

\newcommand{\be}{\begin{equation}}
\newcommand{\ee}{\end{equation}}
\newcommand{\bea}{\begin{eqnarray}}
\newcommand{\eea}{\end{eqnarray}}

\newcommand{\nn}{\nonumber\\}

\newcommand{\UNIT}[1]{\mbox{$\,{\rm #1}$}}

\newcommand{\fm}{\UNIT{fm}}

\def\bi{\begin{itemize}}
\def\ei{\end{itemize}}

\def\fm3{$\mathrm{fm}^3$}
\def\f3{\mathrm{fm}^3}

\def\be{\begin{equation}}
\def\ee{\end{equation}}

\def\B0bar{$\bar{B^0}$}

\begin{document}

\title{Photon elliptic flow in relativistic heavy-ion collisions: \\
hadronic versus partonic sources}

\author{O.~Linnyk}
\email{Olena.Linnyk@theo.physik.uni-giessen.de}
\affiliation{%
 Institute for Theoretical Physics, %
  Justus Liebig University of Giessen, %
  35392 Giessen, %
  Germany %
}

\author{V.P.~Konchakovski}%
\affiliation{%
 Institute for Theoretical Physics, %
  Justus Liebig University of Giessen, %
  35392 Giessen, %
  Germany %
}

\author{W.~Cassing}
\affiliation{%
 Institute for Theoretical Physics, %
  Justus Liebig University of Giessen, %
  35392 Giessen, %
  Germany %
}

\author{E.~L.~Bratkovskaya}%
\affiliation{%
 Institute for Theoretical Physics, %
 Johann Wolfgang Goethe University, %
 60438 Frankfurt am Main, %
 Germany; %
Frankfurt Institute for Advanced Studies, %
 60438 Frankfurt am Main, %
 Germany; %
}

\date{\today}

\begin{abstract}
We study the transverse momentum spectrum and the elliptic flow
$v_2$ of photons produced in Au+Au collisions at
$\sqrt{s_{NN}}=200$~GeV using the Parton-Hadron-String Dynamics
(PHSD) transport approach. As sources for photon production, we
incorporate the interactions of off-shell quarks and gluons in the
strongly interacting quark-gluon plasma (sQGP) ($q+\bar q\to
g+\gamma$ and
 $q(\bar q)+g\to q(\bar q)+\gamma$), the decays of hadrons
($\pi\to\gamma+\gamma$, $\eta\to\gamma+\gamma$,
$\omega\to\pi+\gamma$, $\eta'\to\rho+\gamma$, $\phi\to\eta+\gamma$,
$a_1\to\pi+\gamma$) as well as their interactions
($\pi+\pi\to\rho+\gamma$, $\rho+\pi\to\pi+\gamma$, meson-meson
bremsstrahlung $m+m\to m+m+\gamma$).
The PHSD calculations reproduce the transverse momentum spectrum,
the effective temperature $T_{eff}$ and the elliptic flow $v_2$ of
both inclusive and direct photons as measured by the PHENIX
Collaboration. The photons produced in the QGP contribute slightly
less then 50\% to the observed spectrum, but have small $v_2$. We
find that the large direct photon $v_2$ -- comparable to that of
hadrons -- can be attributed to the intermediate hadronic scattering
channels not subtracted from the data when following the same
extraction procedure for $v_2$ as in the PHENIX experiment. On the
other hand the $v_2$ of direct photons -- as evaluated by the
weighted average of direct photon channels -- gives a lower signal.
The difference between the two extraction procedures for the direct
photon $v_2$ can be attributed to different definitions for the
yield ratio of direct photons to the background photons. The QGP
phase causes the strong elliptic flow of photons indirectly, by
enhancing the $v_2$ of final hadrons due to the partonic interaction
in terms of explicit parton collisions and the mean-field
potentials. We also show that the presence of the QGP radiation is
manifest in the slope of the direct photon spectrum, leading to an
effective slope parameter $T_{eff}$ far above the critical
temperature for the deconfinement phase transition.
\end{abstract}

\pacs{25.75.-q, 13.85.Qk, 24.85.+p}

\maketitle

\section{Introduction}
The electromagnetic emissivity of strongly interacting matter at
finite temperature and baryonic chemical potential is a subject of
longstanding interest and is explored in particular in relativistic
nucleus-nucleus collisions, where the photons (and dileptons)
measured experimentally provide a time-integrated picture of the
collision dynamics. This 'camera' also records the early stages of
such collisions due to the low final state interactions of
electromagnetic signals~\cite{Peitzmann:2001mz}, but the 'picture'
is blurred by the emission at later stages. Fortunately, the 'final
picture' of the hot and dense matter created early in the collision
in part can be restored by independently measuring hadronic channels
and subtracting the associated light signals. The corrected spectra
then are denoted as 'direct photons'.

The recent observation by the PHENIX Collaboration~\cite{PHENIX1}
that the elliptic flow $v_2(p_T)$ of 'direct photons' produced in
minimal bias Au+Au collisions at $\sqrt{s_{NN}}=200$~GeV is
comparable to that of the produced pions was a surprise and in
contrast to the theoretical expectations and
predictions~\cite{Chatterjee:2005de,Liu:2009kq,Dion:2011vd,Dion:2011pp,Chatterjee:2013naa}.
Indeed, the photons produced by partonic interactions in the
quark-gluon plasma phase have not been expected to show considerable
flow because - in a hydrodynamical picture - they are dominated by
the emission at high temperatures, i.e. in the initial phase before
the elliptic flow fully develops. On the other hand, the dominant
hadronic sources of photon production -- decays of $\pi$ and $\eta$
mesons -- have been subtracted by the PHENIX Collaboration from the
total photon spectrum using a model-independent
method~\cite{PHENIX1} and therefore do not explain the observed
strong momentum anisotropy of the direct photons. This has lead also
to the suggestion that the photon $v_2$ observed might be a
signature for more unconventional sources such as the
pre-equilibrium gluon interaction with the magnetic
field~\cite{Bzdak:2012fr,Basar:2012bp}, enhanced emission of photons
at the QGP surface~\cite{Goloviznin:2012dy} or novel assumptions for
the transverse parton acceleration in the
QGP~\cite{Pantuev:2011yh,vanHees:2011vb}.

In the present study we apply the Parton-Hadron-String Dynamics
(PHSD) transport approach to investigate the photon production in
Au+Au collisions at $\sqrt{s_{NN}}=200$~GeV. In the past the PHSD
approach has provided a consistent description of the bulk
properties of heavy-ion collisions -- rapidity spectra, transverse
mass distributions, azimuthal asymmetries of various particle
species -- from low Super-Proton-Synchrotron (SPS) to top
Relativistic-Heavy-Ion-Collider (RHIC) energies~\cite{PHSDqscaling} and
was successfully used also for the analysis of dilepton production
from hadronic and partonic sources at SPS, RHIC and Large-Hadron-Collider (LHC) energies
~\cite{Linnyk:2011hz,Linnyk:2011vx,Linnyk:2012pu}. In the hadronic
sector, PHSD is equivalent to the Hadron-String Dynamics (HSD)
approach~\cite{Cass99,Brat97,Ehehalt}, in which the photon
production at lower SPS energies has been investigated in
Ref.~\cite{ElenaKiselev} with an emphasis on the role of meson-meson
interactions. In the present study we extend the approach to higher
collision energies by explicitly incorporating photon production in
the strongly interacting quark-gluon plasma (sQGP). Indeed, the
deconfined state of matter was found to be  created in heavy-ion
collisions at RHIC~\cite{STARQGP,PHENIXQGP,BRAHMS,PHOBOS} for a
couple of fm/c~\cite{BrCa11} leaving substantial traces especially
in the dilepton yield above invariant masses of 1.2
GeV~\cite{Linnyk:2011hz,Linnyk:2011vx,Linnyk:2012pu}.

The photon radiation from  the partonic phase is consequently
expected to show a large contribution to the transverse momentum
spectrum of produced
photons~\cite{Chatterjee:2005de,Liu:2009kq,Dion:2011vd,Dion:2011pp,Feinberg:1976ua,Shuryak:1978ij,Kapusta:1991qp,Alam:2000bu,
Steffen:2001pv,Srivastava:2000pv,Huovinen:2001wx,Turbide:2003si,d'Enterria:2005vz,Liu:2008eh,Turbide:2007mi},
the slope of which was even used to deduce an 'average temperature'
of the QGP~\cite{PHENIXlast,Adare:2008ab} which we address as an
'apparent inverse slope parameter' or energy scale for the photonic
radiation. The transition to the strongly interacting QGP in the
initial phase of the heavy-ion collisions and the subsequent
hadronization is treated dynamically in the PHSD approach. It is
therefore of interest to calculate the photon production in
relativistic heavy-ion collisions from  hadronic and partonic
interactions consistently within the PHSD transport approach, in
which the microscopic and non-equilibrium evolution of the
nucleus-nucleus collision is independently controlled by a multitude
of other hadronic and electromagnetic observables in a wide energy
range~\cite{BrCa11,Linnyk:2011hz,Linnyk:2011vx,Linnyk:2012pu,Bratkovskaya:2008bf,CasBrat}.

\section{Photon production sources within the PHSD transport approach}
\label{section_phsd}

The {PHSD} model~\cite{CasBrat,BrCa11} is an off-shell transport
approach that consistently describes the full evolution of a
relativistic heavy-ion collision from the initial hard scatterings
and string formation through the dynamical deconfinement phase
transition to the quark-gluon plasma as well as hadronization and to
the subsequent interactions in the hadronic phase. In PHSD the
transition from the partonic to hadronic degrees of freedom is
described by covariant transition rates for the fusion of
quark-antiquark pairs to mesonic resonances or three quarks
(antiquarks) to baryonic states, i.e., by {the} dynamical
hadronization~\cite{Cass08}. The two-particle correlations
{resulting from the finite width of the parton spectral functions}
are taken into account dynamically {in the PHSD} by means of the
{\em generalized} off-shell transport equations~\cite{Cass_off1}
that go beyond the mean field or Boltzmann
approximation~\cite{Cassing:2008nn,Linnyk:2011ee}. The transport
theoretical description of quarks and gluons in the PHSD is based on
the Dynamical Quasi-Particle Model (DQPM) for partons that is
{constructed} to reproduce lattice QCD (lQCD) results for a
quark-gluon plasma in thermodynamic equilibrium. The DQPM provides
the mean-fields for gluons/quarks and their effective 2-body
interactions that are implemented in the PHSD. For details about the
DQPM model and the off-shell transport approach we refer the reader to the
review {in} Ref.~\cite{Cassing:2008nn}.

\subsection{Hadronic sources of photon production}

As sources of photon production - on top of the general dynamical
evolution - we consider hadronic~\cite{Turbide:2003si} as well as
partonic~\cite{Feinberg:1976ua,Shuryak:1978ij} interactions. Let us
first describe the hadronic contributions, which consist of the
photon production from hadronic decays  and the interactions of
final as well as intermediate mesons produced throughout the
evolution of the nucleus-nucleus collision:

1) Photon production by mesonic decays ($\pi^0, \eta, \eta^\prime,
\omega, \phi, a_1$), where the mesons are produced first in
baryon-baryon ($BB$), meson-baryon ($mB$) or meson-meson ($mm$)
collisions. The photon production from the mesonic decays represents
a 'background' for the search of the direct photons, however, this
background can only partly be subtracted from the signal by data
driven methods. Just as in the earlier work within the HSD
model~\cite{ElenaKiselev}, we consider the contributions from the
photon decay of the following mesons:
\begin{eqnarray}
&&\pi^0 \to \gamma+ \gamma,          \nonumber \\
&& \eta \to \gamma + \gamma,         \nonumber\\
&&\eta^\prime \to \rho + \gamma,    \nonumber\\
&&\omega \to \pi^0 + \gamma,        \nonumber\\
&& \phi \to \eta + \gamma,           \nonumber\\
&& a_1 \to \pi + \gamma.             \nonumber
\label{l}\end{eqnarray}
The decay probability is calculated according to the corresponding
branching ratios taken from the latest compilation by the Particle
Data Group~\cite{PDG}, updating slightly the values applied in
earlier HSD investigations at SPS energies~\cite{ElenaKiselev}. The
broad resonances -- including the $a_1, \rho, \omega$ mesons -- in
the initial or final state are treated in PHSD in line with their
(in-medium) spectral functions, as implemented and explained in
detail in Ref.~\cite{ElenaKiselev}.

2) Additionally to the resonance/meson decay channels, the photons can
be produced in mesonic collisions. We consider the direct photon
production  in the scattering processes
\begin{eqnarray}
\pi \pi \rightarrow \rho \gamma , \nonumber\\
\pi \rho \rightarrow \pi \gamma ,  \nonumber
\end{eqnarray}
accounting for all possible charge combinations.

We calculate the cross sections for the processes $\pi \pi
\rightarrow \rho \gamma , \pi \rho \rightarrow \pi \gamma $ as in
Ref.~\cite{ElenaKiselev}, i.e. the total cross section
$\sigma_{\pi\pi \rightarrow \rho\gamma }(s,\rho_N)$ is obtained by
folding the vacuum cross section $\sigma_{\pi\pi \rightarrow
\rho\gamma }^{0}(s,M)$ with the (in-medium) spectral function of the
$\rho$ meson:
\begin{eqnarray}
&&\sigma_{\pi\pi \rightarrow \rho\gamma }(s,\rho_N) =
\label{xs_pipiVVtot}\\
&& \int\limits_{M_{min}}^{M_{max}} dM \
\sigma_{\pi\pi \rightarrow \rho\gamma }^{0}(s,M) \
A(M,\rho_N)\ P(s).\nonumber
\end{eqnarray}
Here $A(M,\rho_N)$ denotes the meson spectral function
for given total width $\Gamma_V^*$:
\begin{eqnarray}
&&A_V(M,\rho_N) = \label{spfunV} \\
&& C_1 \ \frac{2}{\pi} \ \frac{M^2 \ \Gamma_V^*(M,\rho_N)}
{(M^2-M_{0}^{*^2}(\rho_N))^2 + (M {\Gamma_V^*(M,\rho_N)})^2},\nonumber
\end{eqnarray}
with the normalization condition for any $\rho_N$, $
\int_{M_{min}}^{M_{lim}} A_V(M,\rho_N) \ dM =1$, where
$M_{lim}=2$~GeV is chosen as an upper limit for the numerical
integration while the lower limit of the vacuum $\rho$ spectral
function corresponds to the $2\pi$ decay $M_{min}=2 m_\pi$ in vacuum
and $2 m_e$ in medium. $M_0^*$ is the pole mass of the vector meson
spectral function which is $M_0^*(\rho_N=0)=M_0$ in vacuum, however,
might be shifted in the medium (e.g. for the dropping mass
scenario). Furthermore, the vector meson width is the sum of the
vacuum total decay width and collisional width:
\begin{eqnarray}
\Gamma^*_V(M,\rho_N)=\Gamma_V(M) + \Gamma_{coll}(M,\rho_N) . \label{gammas}
\end{eqnarray}
In Eq. (\ref{xs_pipiVVtot}) the function $P(S)$ accounts for the
fraction of the available part of the full spectral function
$A(M,\rho_N)$ at given energy $\sqrt{s}$, integrated over the mass
$M$ up to $M_{max}=\sqrt{s}$, with respect to the total phase space.

The cross section $\sigma_{\pi\pi \rightarrow \rho\gamma }^{0}(s,M)$ is taken from
the model by Kapusta et al. \cite{Kapusta:1991qp} with the
$\rho$-meson mass considered as a dynamical variable, i.e $m_\rho\to M$:
\begin{eqnarray}
 \frac{d \sigma}{dt}\left( \pi^\pm \pi^0 \to \rho ^\pm
\gamma \right) & = & \nn && \hspace{-3.2cm} -\frac{\alpha
g_\rho^2}{16sp_{CM}^2} \left[
\frac{(s-2M^2)(t-m_\pi^2)^2}{M^2(s-M^2)^2}
+\frac{m_\pi^2}{M^2}-\frac{9}{2} \right. \nn && \hspace{-2.4cm}
+\frac{(s-6M^2)(t-m_\pi^2)}{M^2(s-M^2)} +\frac{4
(M^2-4m_\pi^2)s}{(s-M^2)^2} \nn && \hspace{-2.4cm}
\left. +\frac{4(M^2-4m_\pi^2)}{t-m_\pi^2}\left(
\frac{s}{s-M^2}+\frac{m_\pi^2}{t-m_\pi^2} \right) \right].
\end{eqnarray}
The photon production in the $\pi+\rho$ interaction is calculated
analogously (cf. Ref.~\cite{ElenaKiselev} for details).

3) Colliding charged mesons can also radiate photons by the
bremsstrahlung process $m+m\to m +m + \gamma$. The implementation of
photon bremsstrahlung from hadronic reactions in transport
approaches is based on the 'soft photon' approximation. The
soft-photon approximation (SPA) \cite{Gale87} relies on the
assumption that the radiation from internal lines is negligible and
the strong interaction vertex is on-shell. In this case the strong
interaction part and the electromagnetic part can be separated, so
the soft-photon cross section for the reaction $m_1 + m_2 \to m_1 +
m_2 + \gamma$ can be written as
\begin{eqnarray}
q_0\frac{d^3 \sigma^\gamma}{d^3 q} & = & \frac{\alpha}{4 \pi}
\frac{{\bar \sigma (s)}}{q_0^2},
            \label{brems} \\
{\bar \sigma(s)} & = & \frac{s - (M_1 + M_2)^2}{2 M_1^2} \sigma(s),
\nonumber
\end{eqnarray}
where $M_1$ is the mass of the charged accelerated particle, $M_2$
is the mass of the second particle; $q_0, q$ are the energy and
momentum of the photon. In (\ref{brems})  $\sigma(s)$ is the
on-shell elastic cross section for the reaction $m_1+m_2\to
m_1+m_2$.
The photon production via the meson-meson bremsstrahlung process in
lower energy heavy-ion collisions was calculated in exactly this
fashion in the earlier work within the hadronic HSD transport
model~\cite{ElenaKiselev}. Here, we will calculate the photon
bremsstrahlung from all elastic meson-meson scatterings $m_1 + m_2$,
including now also the vector mesons ($m_i = \pi, \eta, K, \bar K,
K^0, K^*, \bar K^*, K^{*0}, \eta', \omega, \rho, \phi, a_1$), which
occur during the heavy-ion collisions by applying the SPA formula
(\ref{brems}). Let us point out that the resulting yield of the
bremsstrahlung photons depends on the model assumptions such as the
cross section for the meson-meson elastic scattering (we assume 10
mb for all meson species), incoherence of the individual scatterings
and the soft photon approximation. The adequacy of the SPA
assumption has been shown in Ref.~\cite{PhysRevD.53.4822}, however,
a possible theoretical uncertainty of up to a factor of 2 has to be
kept in mind.

The PHSD and HSD are off-shell transport models and thus allow to
study the effect of the vector meson spectral function modification.
In particular the photon production in secondary meson interactions
is sensitive to the properties of the vector mesons in the
medium~\cite{ElenaKiselev,Song:1994zs,Song:1993ae}. In this respect,
we stress here that the yields and the in-medium spectral functions
of vector mesons in PHSD have been independently constrained by the
comparison to the data on dilepton mass-spectra (see
Refs.~\cite{Linnyk:2011hz,Linnyk:2011vx,Bratkovskaya:2008bf}).

\subsection{Partonic sources of photon production}

\begin{figure}
\includegraphics[width=0.49\textwidth]{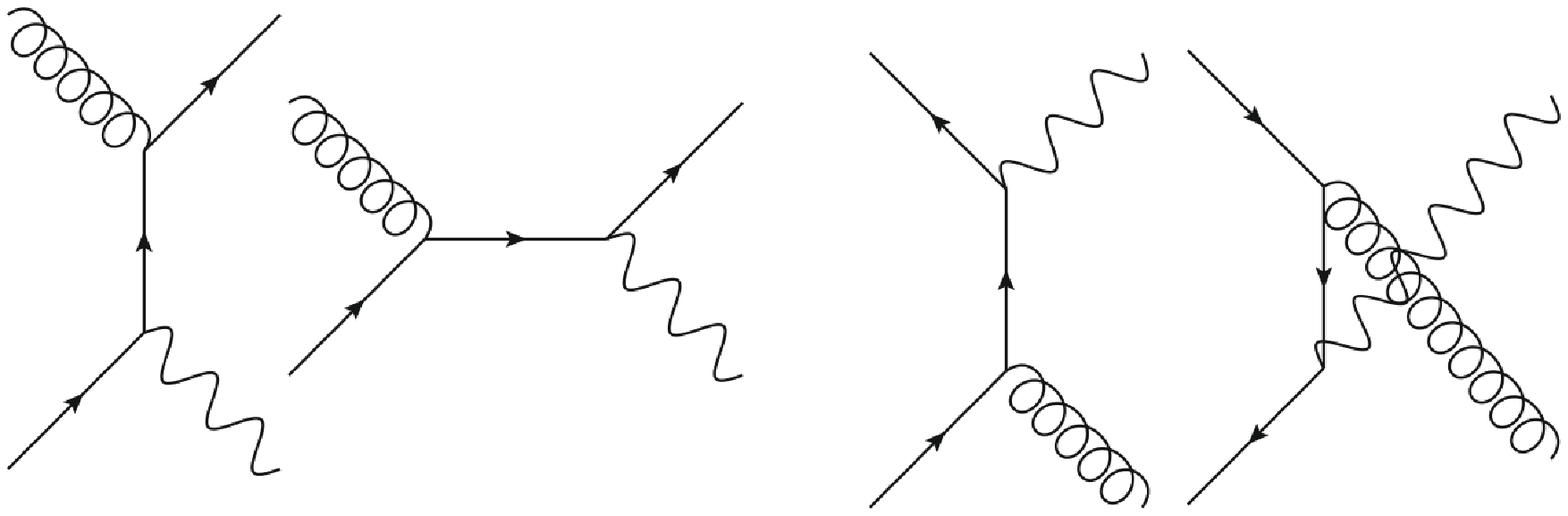}
\caption{(Color on-line) Feynman diagrams for the partonic sources
($q({\bar q})+g \rightarrow q({\bar q})+\gamma$ and $q+{\bar q}
\rightarrow g+\gamma$) included in the PHSD calculations.}
\label{fig0}
\end{figure}

We continue with the description of photon production in the
interactions of quarks and gluons in the quark-gluon plasma, which
dominantly proceeds through the quark annihilation and the gluon
Compton scattering processes:
\begin{eqnarray}
q+{\bar q} & \rightarrow & g+\gamma \nn
q({\bar q})+g & \rightarrow & q({\bar q})+\gamma. \nonumber
\end{eqnarray}
The diagrams contributing to these scattering processes at tree
level are presented in Fig.~\ref{fig0}.

In the strongly interacting QGP, the gluon and quark propagators (in
PHSD) differ significantly from the non-interacting propagators such
that bare production amplitudes can no longer be used. The off-shell
quarks and gluons have finite masses and widths, which parametrize
the resummed interaction of the QGP constituents. The perturbative
QCD results for the cross sections of the processes in
Fig.~\ref{fig0} have to be generalized in order to include the
finite masses for fermions and gluons as well as their broad
spectral functions. In Refs.~\cite{Marzani:2008uh}, the influence of
the gluon off-shellness (fixed to $m_g^2=|\vec{k_g}|^2$) on the
photon production was studied but the quark masses had been
neglected and the spectral functions were assumed to be
$\delta$-functions (quasi-particle approximation). On the other
hand, in Ref.~\cite{Wong:1998pq} a finite quark mass was
incorporated in the elementary cross sections for both the quark
annihilation and the gluon-Compton scattering processes (though the
gluon was taken to be massless and the quasiparticle approximation
remained). The resulting cross sections of Ref.~\cite{Wong:1998pq}
are instructive and still compact enough to be explicitly shown here
for illustration of the quark mass effect:
\begin{eqnarray}
\hspace{-0.5cm} \frac{d \sigma }{d t} (q\bar q\to\gamma g) &=& \nn
&& \hspace{-2.5cm} - \frac{4}{9} \left( \frac{e_q^2}{e^2} \right)^2
\frac{8 \pi \alpha_S \ \alpha_{EM}}{s (s-4 m_q^2)} \left[  \left(
\frac{m_q^2}{t-m_q^2} + \frac{m_q^2}{u-m_q^2} \right)^2 \right. \nn
&& \hspace{-2.5cm} \left. + \left( \frac{m_q^2}{t-m_q^2} +
\frac{m_q^2}{u-m_q^2} \right) - \frac{1}{4}\left(
\frac{t-m_q^2}{u-m_q^2} + \frac{u-m_q^2}{t-m_q^2} \right) \right]
\label{pQCD1}
\end{eqnarray}
\begin{eqnarray}
\hspace{-0.5cm} \frac{d \sigma }{d t} (g q\to\gamma q) &=& \nn
&& \hspace{-2.5cm} \ \  \frac{1}{6} \left( \frac{e_q^2}{e^2}
\right)^2 \frac{8 \pi \alpha_S \ \alpha_{EM}}{(s- m_q^2)^2} \left[
\left( \frac{m_q^2}{s-m_q^2} + \frac{m_q^2}{u-m_q^2} \right)^2
\right. \nn
&& \hspace{-2.5cm} \left. + \left( \frac{m_q^2}{s-m_q^2} +
\frac{m_q^2}{u-m_q^2} \right) - \frac{1}{4}\left(
\frac{s-m_q^2}{u-m_q^2} + \frac{u-m_q^2}{s-m_q^2} \right) \right]
\label{pQCD2}
\end{eqnarray}
It is obvious from equations (\ref{pQCD1}) and (\ref{pQCD2}) that
the quark off-shellness leads to higher twist corrections ($\sim
m_q^2/s,m_q^2/t,m_q^2/u$). These corrections are small in hard
hadron scattering at high center-of-mass energy $\sqrt{s}>10$~GeV,
but become substantial for  photon production in the sQGP, where the
characteristic $\sqrt{s}$ of parton collisions is of the order of a
few GeV.

However, the formulae (\ref{pQCD1}) and (\ref{pQCD2}) have to be
further generalized in order to be included into the PHSD transport
approach: we have to incorporate the finite mass for the gluon and
departe from the quasiparticle approximation. Note also that in
(\ref{pQCD1}) and (\ref{pQCD2}) the masses of quarks and antiquarks
were assumed equal, which is not the case for the offshellnesses of
the broad strongly-interacting particles in the PHSD. The evaluation
of cross sections for dilepton production by off-shell partons,
taking into account finite masses for quarks, antiquarks (with
generally $m_q\ne m_{\bar q}$) and gluons $m_g$ as well as their
finite spectral width (by integrating over the mass distributions in
analogy to equation (\ref{xs_pipiVVtot})) has been carried out in
Refs.~\cite{Linnyk:2004mt,olena2010} We refer the reader to the
work~\cite{olena2010} for details of the calculations, where the
phenomenological parametrizations from the DQPM for the quark and
gluon propagators and their interaction strength were used. Since
the resulting formulae are quite lengthy, we do not repeat them
here. In order to obtain now the cross sections for the {\em real}
photon production, we use the relation between the real photon
production cross section and the cross section for dilepton
production~\cite{PHENIXlast}:
\begin{equation}
\label{realvirtual} \frac{d\sigma (\gamma)}{dt}=\lim_{M\to 0}
\frac{3\pi}{\alpha} \frac{M^2}{L(M)} \frac{d^2 \sigma (e^+e^-)}{dM^2
dt},
\end{equation}
where $M^2$ is the invariant mass squared of the lepton pair, i.e.
the virtuality of the virtual photon, and the kinematical factor
$L(M)$ is given by
\begin{equation}
L(M)=\sqrt{1-\frac{4 m_e^2}{M^2}}(1+\frac{2m_e^2}{M^2}),
\end{equation}
with $m_e$ denoting the lepton mass.

We take $d^2\sigma (e^+e^-)/dM^2 dt$ from Ref.~\cite{olena2010} and
use relation (\ref{realvirtual}) to implement the real photon
production in the off-shell quark and gluon interactions into the
PHSD transport approach. In each interaction of $q+\bar q$ or
$q/\bar q+g$, the photon production probability and the elliptic
flow of the produced photon are recorded differentially in
transverse momentum $p_T$ and rapidity $y$.

\begin{figure}
\includegraphics[width=0.49\textwidth]{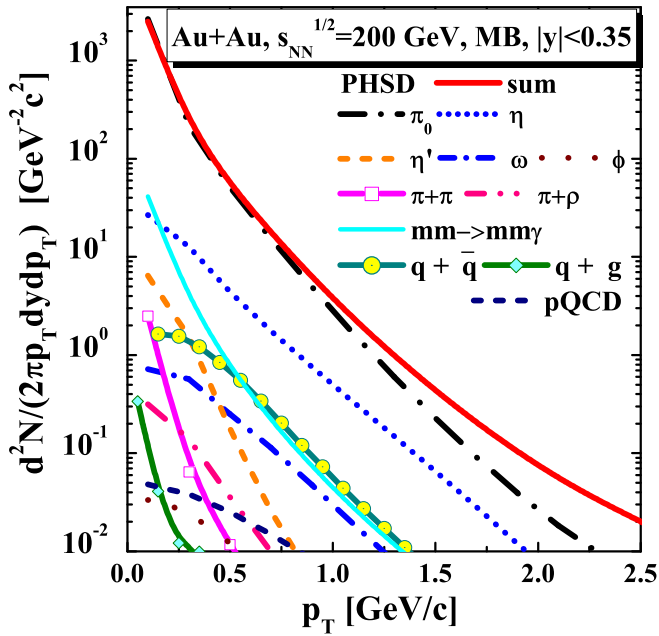}
\caption{(Color on-line) The channel decomposition of the inclusive
photon transverse momentum ($p_T$) spectrum from PHSD  for minimal
bias Au+Au collisions at $\sqrt{s_{NN}}=200$~GeV (full solid upper
line) at mid-rapidity $|y|< 0.35$. Channel description is given in
the text and the legend.} \label{incl_spectra}
\end{figure}

\section{Results}
\label{sectionresults}

\subsection{Spectra}

\begin{figure}
\subfigure{ 
\resizebox{0.48\textwidth}{!}{%
 \includegraphics{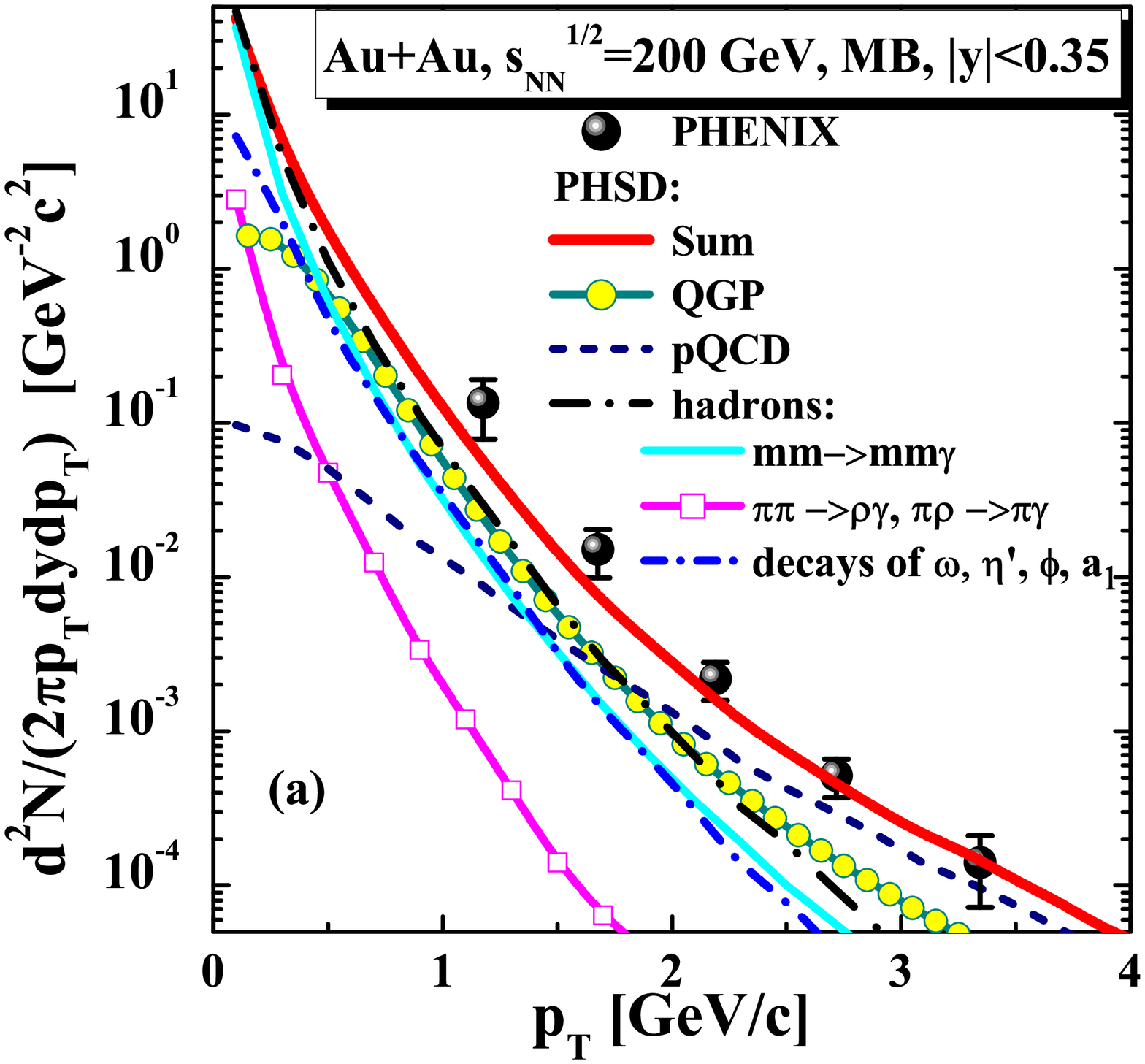}
} } \subfigure{ 
\resizebox{0.48\textwidth}{!}{%
 \includegraphics{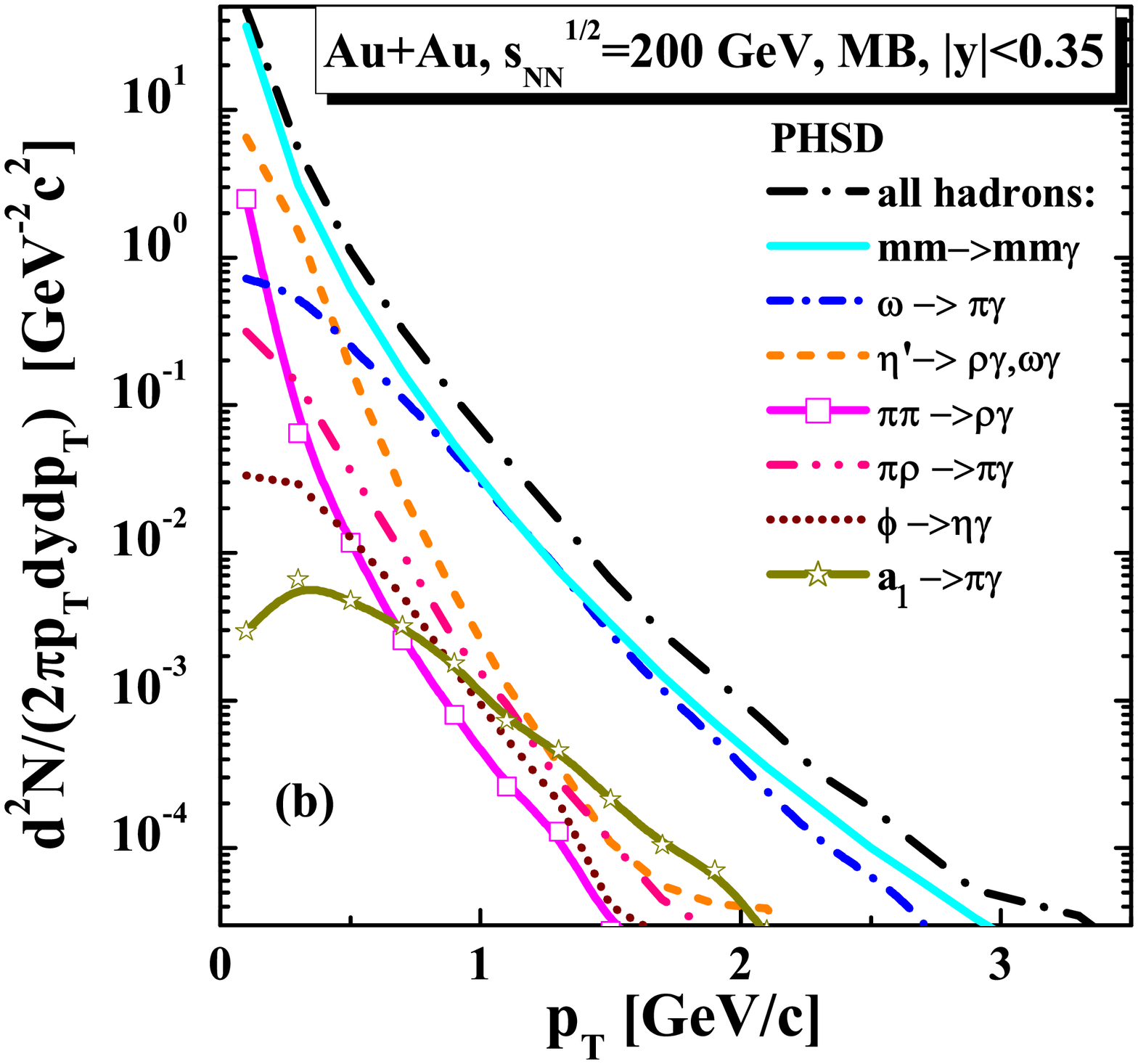}
} } \caption{(Color online) Top panel (a): Direct photons (sum of
all photon production channels except the $\pi$- and $\eta$-meson
decays) from the PHSD approach (red solid line) in comparison to the
data of the PHENIX
Collaboration~\protect{\cite{PHENIXlast,Adare:2008ab} for minimal
bias collisions of Au+Au at $\sqrt{s_{NN}}=200$~GeV } (black
symbols). The various channels are described in the legend. Bottom
panel (b): Detailed channel decomposition for the direct photon
production by the hadronic scatterings and decays. }
\label{dir_spectra}
\end{figure}

The results for the inclusive photon spectrum as a sum of all the
considered partonic as well as hadronic sources for the photons
produced in  minimal bias Au+Au collisions at
$\sqrt{s_{NN}}=200$~GeV is presented in Fig.~\ref{incl_spectra} as a
function of the transverse momentum $p_T$ at mid-rapidity $|y|<
0.35$. In our calculations of the total photon spectrum the
following sources are taken into account: the decays of $\pi^0$,
$\eta$,  $\omega$, $\eta$', $\phi$ and $a_1$ mesons; the reactions
$\pi+\rho\to\pi+\gamma$, $\pi+\pi\to \rho+\gamma$; the photon
bremsstrahlung in meson-meson collisions $m+m\to m+m+\gamma$; photon
production in the QGP in the processes $q+{\bar q}  \to g+\gamma$,
and $q({\bar q})+g \to q({\bar q})+\gamma$; and the photon
production in the initial hard collisions ("pQCD"), which is given
by the hard photon yield in p+p collisions scaled with $N_{coll}$
(line taken from Ref.~\cite{PHENIXlast,Adare:2008ab}). The leading
contributions are the decays of $\pi^0$ and $\eta$ mesons. Since
these 'late' hadronic sources are less sensitive to the creation of
the hot and dense medium and to its properties, they are subtracted
experimentally to access the `direct' photon spectrum.

\begin{figure}
\resizebox{0.48\textwidth}{!}{%
 \includegraphics{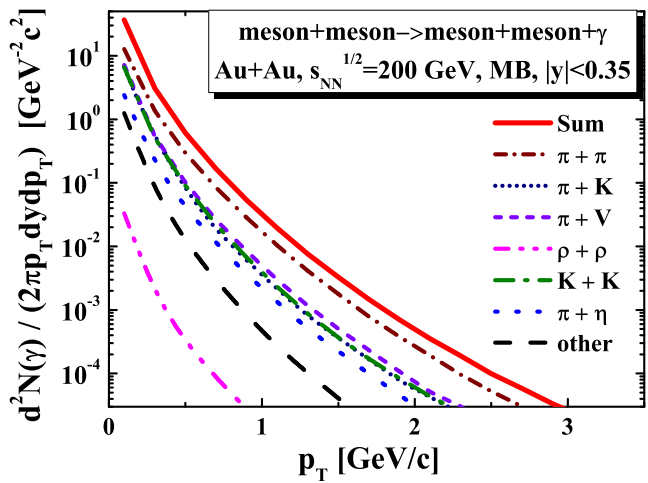}
} \caption{(Color online) The spectra of photons at mid-rapidity
produced in the meson-meson bremsstrahlung process (red solid line)
in minimal bias collisions of Au+Au at $\sqrt{s_{NN}}=200$~GeV
within the PHSD approach. The various colored lines show the
contribution of different mesonic collision channels.}
\label{MMchannels}
\end{figure}

\begin{figure}
\resizebox{0.48\textwidth}{!}{%
 \includegraphics{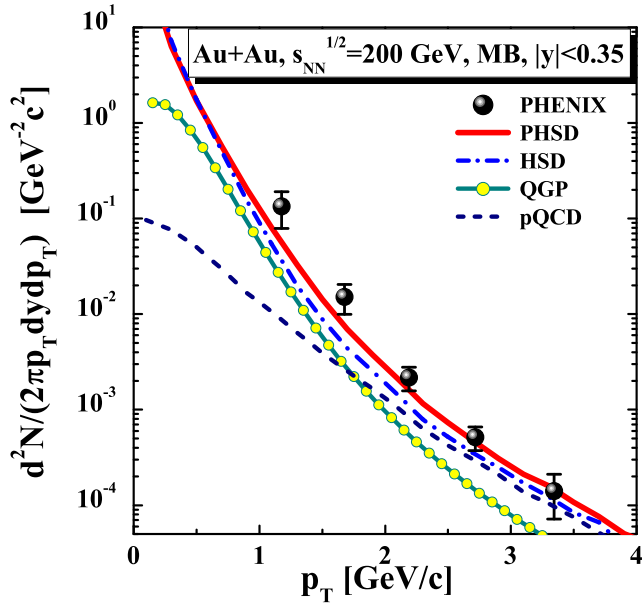}
} \caption{(Color online) Comparison of the direct photons from the
PHSD approach (red solid line) and the hadron/string model HSD
(dash-dotted blue line), the latter having no transition to the
deconfined phase. The data of the PHENIX
Collaboration~\protect{\cite{PHENIXlast,Adare:2008ab}} are shown by
the black symbols. } \label{HSD}
\end{figure}

The experimental data on the remaining 'direct' photon spectrum are
compared to the PHSD calculations (without $\pi^0$ and $\eta$
decays) in Fig.~\ref{dir_spectra}. The measured transverse momentum
spectrum $dN/dp_T$ is reproduced, if the partonic and the remaining
hadronic sources are summed up (upper solid line). We find that the
radiation from the sQGP constitutes slightly less than half of the
observed number of photons. The radiation from hadrons and their
interaction -- which are not measured separately so far -- give a
considerable contribution, too, especially at low transverse
momentum. The dominant hadronic sources are the meson decays and the
meson-meson bremsstrahlung. While the former (e.g. the decays of
$\omega$, $\eta$', $\phi$ and $a_1$ mesons) can be subtracted from
the spectra once the mesonic yields are determined independently by
experiment, the reactions $\pi+\rho\to\pi+\gamma$, $\pi+\pi\to
\rho+\gamma$ and the meson-meson bremsstrahlung $m+m\to m+m+\gamma$
cannot be removed by model-independent methods. In this respect, the
$m+m\to m+m+\gamma$ channel is especially interesting, thus we show
it in more detail in Fig.~\ref{MMchannels}, where the photon
production in the scattering of pions gives the dominant
contribution among the mesonic collision channels.
On the other hand, we explicitly show in Fig.~\ref{HSD} that the
purely hadron/string scenario as incorporated within the HSD
transport approach (dash-dotted line) clearly underestimates the data.

Though the spectrum presented in Fig.~\ref{dir_spectra} is not
thermal, one can attempt to approximate it by an exponential
spectrum for low transverse momenta to extract an effective slope or `effective
temperature' as done by the PHENIX Collaboration in
Ref.~\cite{PHENIXlast,Adare:2008ab}. In practice, we fit the
spectrum in the range $0.3<p_T<3.0$~GeV by the function $A
\exp(-p_T/T_{eff}) + T_{AA} (dN/dp_T)_{pp}$, where the latter term
denotes the scaled photon yield from $p+p$ collisions, shown in
Fig.~\ref{dir_spectra} by the short-dashed blue line and labeled as
"pQCD". In the theoretical approach, we have the possibility to
separate the yields according to the production sources and
therefore to extract the `effective temperatures' of the photons
stemming from parton interactions and those of hadronic origin. The
extracted `effective temperatures' are shown in Table I. One observes
that the characteristic `temperature' scale for the photons produced
in the QGP is higher than for the photons emitted by mesonic decays and
their interactions. The $T_{eff}$ of the total spectrum agrees with
the experiment value within errors.

\begin{table}
\begin{tabular}{|c|c|c||c|}
\hline
  \multicolumn{4}{|c|}{ The slope parameter $T_{eff}$ (in MeV) \phantom{\Large I}} \\
\hline
  \multicolumn{3}{|c||}{ \phantom{\Large I} PHSD \phantom{\Large I} } & \phantom{\Large I} PHENIX \phantom{\Large I} \\
\cline{1-3} \phantom{\Large I} QGP \phantom{\Large I} &
\phantom{\Large I} hadrons \phantom{\Large I} & \phantom{\Large I}
Total \phantom{\Large I} & \protect{\cite{Adare:2008ab}}
\\ \hline \ $260\pm20$ \ & \ $200\pm20$ \ & \ $220\pm20$ \ &
\phantom{\Large I} $233\pm14\pm19$ \
\\ \hline
\end{tabular}
\caption{The slope parameter $T_{eff}$ of the direct photon spectrum
($0.3<p_T<3.0$ GeV/c) in minimal bias Au+Au collisions at
$\sqrt{s_{NN}}=200$~GeV.} \label{table}
\end{table}

\begin{figure}
\resizebox{0.48\textwidth}{!}{%
 \includegraphics{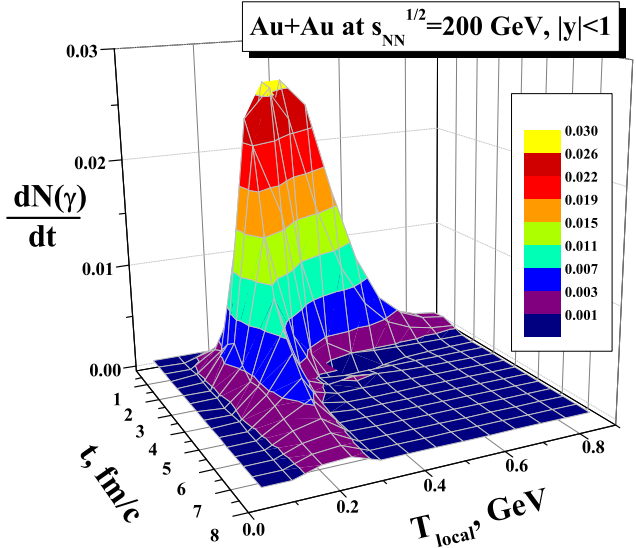}
}
\caption{(Color online) The  photon production rate versus time and
the local 'temperature' at the production point in 
 Au+Au collisions at mid-rapidity.} \label{3D}
\end{figure}

However, the 'effective temperature' discussed above is just the
slope parameter of the transverse momentum spectrum and does not
directly represent the 'temperature' of the QCD medium in which the
photons are produced. We recall that PHSD does not assume global or
local thermal equilibrium, since the produced matter, at least
initially, is far from equilibrium. Consequently, at each time step,
the photons are produced in an ensemble of microscopic cells with
different local energy density, which varies considerably from cell
to cell. We can study the distribution in the local energy density
of photon production in relativistic heavy-ion collisions as a
function of time. The results are presented in Fig.~\ref{3D}, where
for illustration purposes the energy density scale was recalculated
into a temperature scale using the lattice QCD equation of state in
equilibrium from Ref. \cite{Borsanyi:2010bp}. Fig.~\ref{3D} shows
the photon production rate in the QGP at mid-rapidity ($|y|<1$) as
created in heavy-ion collisions at $\sqrt{s_{NN}}=200$~GeV versus
time and the local 'temperature'. It is clearly seen that no
universal 'temperature' can be assigned to the whole volume of the
QGP for fixed times (even in the mid-rapidity region). Instead we
see a broad distribution of 'temperatures', which becomes narrower
with increasing time with the average temperature gradually
decreasing (cooling).

\subsection{Elliptic flow of inclusive photons}

\begin{figure}
\includegraphics[width=0.48\textwidth]{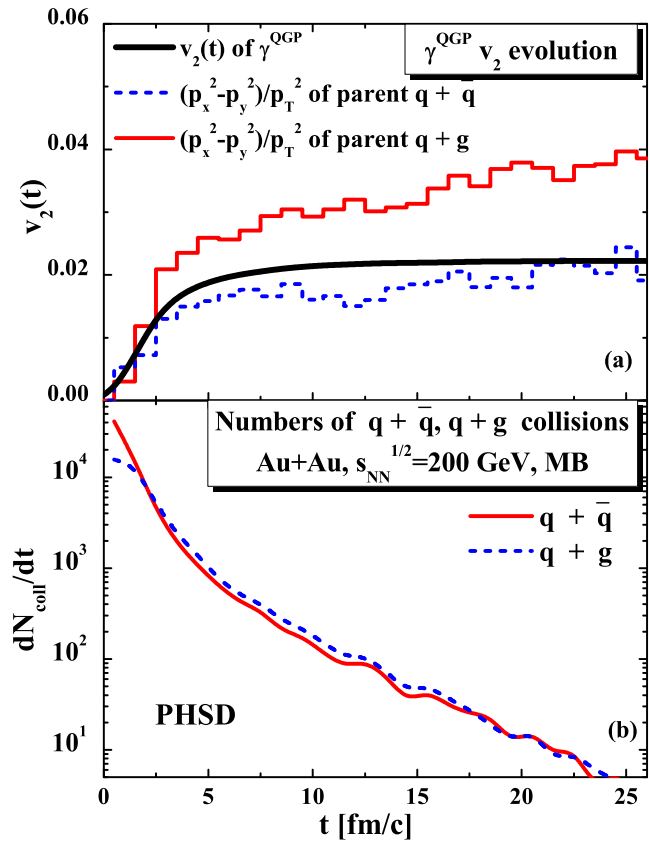}
\caption{(Color on-line) The azimuthal asymmetry of the momentum
distribution for the parent partons - producing photons - and the
time evolution for the elliptic flow $v_2$ of the photons created in
these collisions (top panel). The number of photon-producing
collisions as a function of time (bottom panel) for minimal bias
Au+Au collisions at $\sqrt{s_{NN}}=200$~GeV.} \label{v2_time}
\end{figure}

Since almost half of the direct photons measured by PHENIX stem from
the collisions of quarks and gluons in the deconfined medium created
in the initial phase of the collision, we first investigate the
amount of elliptic flow that is carried by the colliding (`parent')
partons. Note that the cross sections of the $q+\bar q\to g+\gamma$
and $q+g\to q+\gamma$ processes are isotropic in the azimuthal
angle, thus the azimuthal asymmetry of the produced photons is
generated from the direction of the summed initial momenta of the
colliding partons ($\vec p_1, \vec p_2$), i.e. $\vec q=\vec p_1 +
\vec p_2$. In Fig.~\ref{v2_time} we show (top panel) the asymmetry
$(q_X^2-q_Y^2)/(q_X^2+q_Y^2)$ for the total momentum $\vec q$ of the
quark+antiquark (quark+gluon) pairs, which have suffered a collision
and produce a photon through the $q+\bar q\to g+\gamma$ ($q+g\to
q+\gamma$) process. The bottom panel of Fig.~\ref{v2_time} presents
the number of such collisions versus time. We observe that the
parton collisions - producing photons - take place throughout the
evolution of the collision but the collision rate falls rapidly with
time and thus the production of photons from the QGP is dominated by
the early times. As a consequence, the elliptic flow `picked up' by
the photons from the parent parton collisions (given by the black
solid line in the top panel of Fig.~\ref{v2_time}) saturates after
about 5~fm/c and reaches a relatively low value of about $0.02$,
only. In comparison to the elliptic flow of the finally produced
hadrons, the $v_2$ of the QGP photons is lower by about an order of
magnitude.

\begin{figure}
\includegraphics[width=0.48\textwidth]{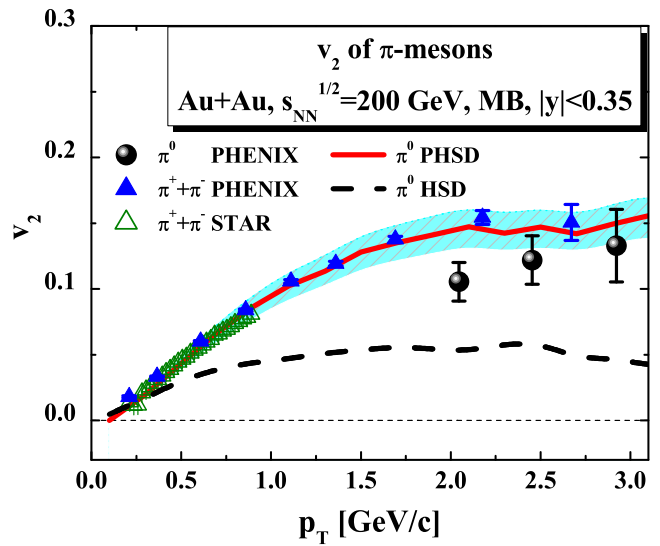}
\caption{(Color on-line) Elliptic flow of neutral pions from the
PHSD approach (red solid line) and the data for the neutral and
charged pion $v_2$~\cite{PHENIX1,Adams:2004bi,Adler:2003kt} for
minimal bias Au+Au collisions at $\sqrt{s_{NN}}=200$~GeV. The result
from HSD is displayed by the dashed line, while the hatched area
denotes the statistical uncertainty in the PHSD.} \label{v2_pi}
\end{figure}

\begin{figure}
\includegraphics[width=0.48\textwidth]{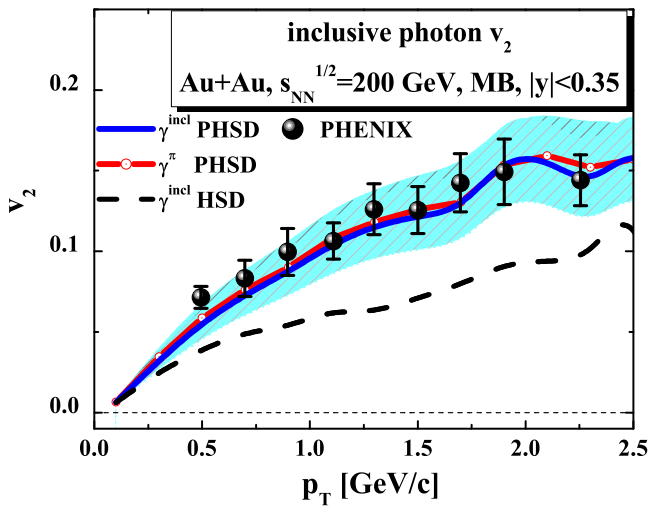}
\caption{(Color on-line) Elliptic flow of inclusive photons from the
PHSD approach  for minimal bias Au+Au collisions at
$\sqrt{s_{NN}}=200$~GeV. The data are from Ref.~\cite{PHENIX1}. The
result from HSD is displayed by the dashed line, while the hatched
area denots the statistical uncertainty in the PHSD.}
\label{v2_incl}
\end{figure}

In Fig.~\ref{v2_pi} we show the $v_2$ of pions as calculated in the
PHSD transport approach in comparison to the data from the PHENIX
and STAR Collaborations~\cite{PHENIX1,Adams:2004bi,Adler:2003kt}.
One can see that the $v_2$ of produced $\pi$-mesons reaches $v_2
\approx$ 0.15 and is well reproduced by the PHSD calculations as a
function of the meson transverse momentum $p_T$. Note that in all
the calculations presented in
Figs.~\ref{v2_pi},\ref{v2_incl},\ref{v2_dir} the reaction plane
correction has been applied event-by-event.
The strong elliptic flow of hadrons for momenta $p_T<3$~GeV has the
following origin: The dynamical hadronization happens at the end of
the QGP evolution, i.e. when the elliptic flow of partons has
already developed. Thus the hadrons pick up the collective
acceleration of the partons. Contrary, photons from partonic
interactions are essentially radiated during the first few fm/c of
the collision dynamics in the central area of high energy density,
while no mesons are produced in this space-time regime.
We recall that the PHSD successfully describes of the total $v_2$ of
final hadrons, while the purely hadronic scenario (HSD) leads to
a substantial underestimation of elliptic flow \cite{PHSDasymmetries},
manly due to the lack of partonic interactions and a repulsive parton
mean-field potential. We confirm this conclusion by presenting the
$p_T$-dependence of the pion elliptic flow from the HSD approach in
Fig.~\ref{v2_pi} for comparison (dashed line).

Next, we present in Fig.~\ref{v2_incl} the elliptic flow of the
inclusive photons produced in  minimal bias Au+Au collisions at
$\sqrt{s_{NN}}=200$~GeV from the PHSD (red solid line) and the
experiment as measured by the PHENIX Collaboration. The statistical
uncertainty of about 10\% is shown explicitly in the plot by the
hatched area. We find that the data are reproduced within errors.
The blue line with symbols shows the calculated $v_2$ of photons
produced in pion decays. As we have seen above, the pion decay
photons dominate the inclusive photon spectrum. Since the elliptic
flow of pions is under control in PHSD (cf. Fig~\ref{v2_pi}), the
total photon $v_2$ is naturally also well described. Again, the HSD
line calculated in the absence of the QGP (dashed line)
underestimates the data by about a factor of 2.

\begin{figure}
\includegraphics[width=0.48\textwidth]{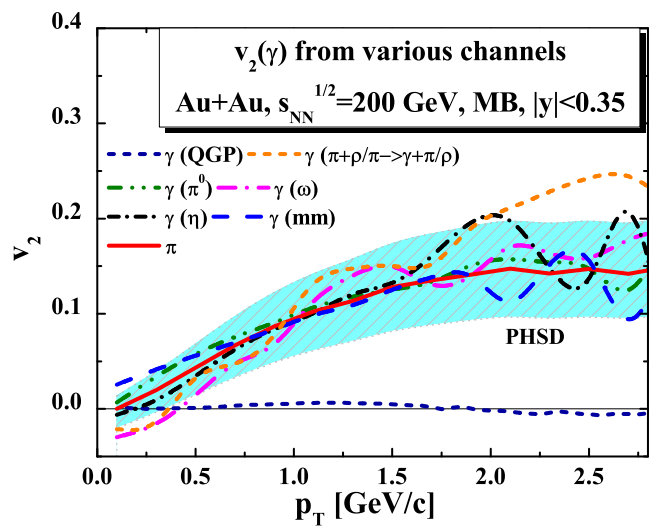}
\caption{(Color on-line) Elliptic flow of photons from different
production channels in minimal bias (MB) Au+Au collisions at
$\sqrt{s_{NN}}=200$~GeV from the PHSD approach.} \label{v2_channels}
\end{figure}

In Fig.~\ref{v2_channels} we show separately the $v_2$ of photons
produced in various channels: QGP (short-dashed blue line), meson
reactions $\pi+\pi\to \rho+\gamma$ and $\pi+\rho\to\pi+\gamma$
(short-dashed orange line), $\pi_0$ decay (dash-dot-doted green
line), $\omega$ decay (dash-dotted magenta line), $\eta$ decay
(dash-dotted black line), and the meson-meson bremsstrahlung $m+m\to
m+m+\gamma$ (dashed blue line). Note that the photons from hadronic
channels roughly have the same $v_2(p_T)$ within errors (indicated
in the plot as the hatched error band) and approximately equal to
the $v_2$ of $\pi$-mesons given for comparison on the same plot (red
line), while the $v_2$ of the photons produced in the QGP is about a
factor $10$ less.

\subsection{Elliptic flow of direct photons}

We have shown above that the PHSD describes the data on the eliptic
flow of inclusive photons produced in $Au+Au$ collisions at
$\sqrt{s}=200$~GeV. In order to obtain the flow of direct photons,
the hadron decay background has to be subtracted from the inclusive
photon flow. This can be done by two procedures which we describe
below. Firstly, the flow of the direct photon channels can be
combined as a weighted sum (procedure 1). In this case, we will take
into account only the binary channels (the parton scatterings in the
QGP as well as hadron reactions). The hadron decay photons will not
enter the sum. The second possibility (procedure 2) is to follow the
same background subtraction procedure as in the experiment (i.e.
estimating the hadron decay contributions and subtracting their flow
from the inclusive photon flow with relative weight). We present the
results of both methods and estimate each method's error in the
following.

\subsubsection{Procedure 1}

\begin{figure}
\includegraphics[width=0.48\textwidth]{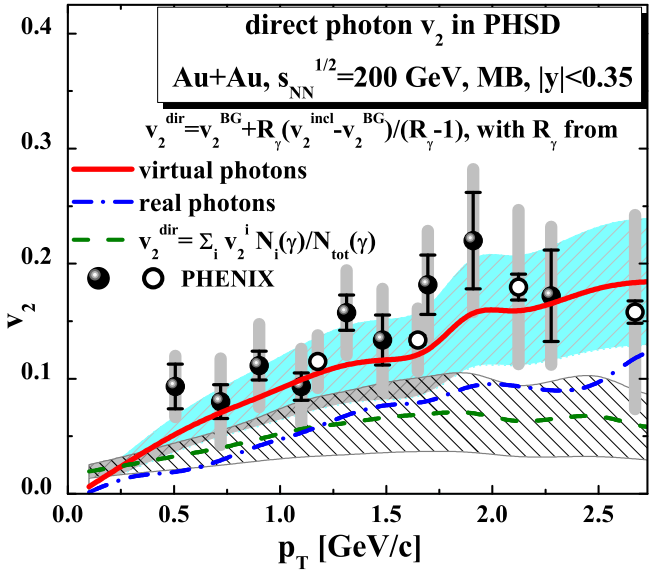}
\caption{(Color on-line) Elliptic flow of direct photons (hadron
decays excluded) in the PHSD approach for minimal bias Au+Au
collisions at $\sqrt{s_{NN}}=200$~GeV. The data are from
Refs.~\cite{PHENIX1,Tserruya:2012jb}. The results from the PHSD are
displayed by the solid red line, equation (\protect\ref{dir1}), and
by the dash-dot blue line, by applying Eq.~(\protect\ref{dir2}).}
\label{v2_dir}
\end{figure}

We can calculate the direct photon $v_2$ (in PHSD) by summing up the
elliptic flow of the individual channels contributing to the direct
photons, using their contributions to the spectrum as the relative
$p_T$-dependent weights, $w_i(p_T)$, i.e.
\bea \label{dir2} &  v_2 (\gamma^{dir}) = \sum _i  v_2 (\gamma^{i})
w_i (p_T) =  \frac{\sum _i  v_2 (\gamma^{i}) N_i (p_T)}{\sum_i N_i
(p_T)}, & \! \! \\
%
& \! \! i=(q\bar q \! \to \! g \gamma, q g \! \to \! q \gamma, \pi
\pi/\rho \! \to \! \rho/\pi \gamma, m m \! \to \! m m \gamma,
\mbox{pQCD}). & \nonumber \eea
The index $i$ denotes only the binary channels, both the partonic
quark-gluon interaction channels and the meson reactions which
cannot be separated presently experimentally by model-independent
methods: $q\bar q \! \to \! g \gamma$, $q/\bar q g \! \to \! q/\bar
q \gamma$, $\pi+\pi\to \rho+\gamma$, $\pi+\rho\to\pi+\gamma$,
$m+m\to m+m+\gamma$ and the hard photons as produced in the $p+p$
collisions scaled with $N_{coll}$ (``pQCD"). This procedure is
possible in our approach, since we model each contributing channel
and calculate their spectra and $v_2$ individually as a function of
$p_T$. For the pQCD photons from the initial hard nucleon-nucleon
collisions we assumed zero $v_2$. The direct photon elliptic flow
calculated in this way is presented in Fig.~\ref{v2_dir} by the
dashed green line. The analysis of the statistical error for this
formula gives
\be \! \delta v_2(\gamma^{dir}) =  \sum_i \delta v_2 (\gamma^{i})
\frac{N_i(p_T)}{N(p_T)} + \sum_i \delta N_i(p_T)
\frac{v_2(\gamma^{i})}{N(p_T)} \!\! \ee
and is shown by the lower hashed error band in Fig.~\ref{v2_dir}.
%

\subsubsection{Procedure 2}

The experimental collaboration has extracted the elliptic flow of
direct photons $v_2(\gamma^{dir})$ from the measured inclusive
photon $v_2(\gamma^{incl})$ by subtracting the -- partly
extrapolated --  hadron decay sources ($\pi_0$, $\eta$, $\omega$,
$\eta'$, $\phi$, $a_1$) as follows~\cite{PHENIX1,Gaboprivat}:
%
%
\begin{eqnarray} \label{dir1}
\hspace{-0.5cm} v_2 (\gamma^{dir}) & \! = \! & \frac{R_\gamma
v_2(\gamma^{incl}) - v_2 (\gamma^{BG})}{R_\gamma -1} \nn & \! = \! &
v_2 (\gamma^{BG}) + \frac{R_\gamma }{R_\gamma -1} (
v_2(\gamma^{incl})-v_2(\gamma^{BG}))
\end{eqnarray}
where $$R_\gamma=N^{incl}/N^{BG}$$ denotes the ratio of the
inclusive photon yield to that of the ``background" (i.e. the
photons stemming from the decays of $\pi_0$, $\eta$, $\omega$,
$\eta'$, $\phi$ and $a_1$ mesons), $v_2(\gamma^{BG})$ in the
elliptic flow of the background photons. The PHENIX collaboration
has used the cocktail of mesons and a scaling assumption for their
respective $v_2$ to extrapolate the background $v_2(\gamma^{BG})$.
In Fig.~\ref{v2_channels} we have shown the $v_2$ for leading
individual channel separately from the PHSD; one can see that
$v_2(\gamma^{BG})\approx v_2(\pi)$.
%

\begin{figure}
\resizebox{0.48\textwidth}{!}{%
 \includegraphics{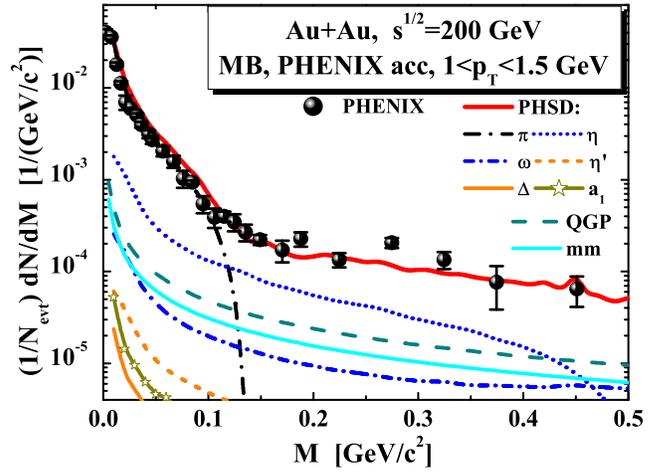}
} \caption{(Color online) The invariant mass spectrum of dileptons
at high transverse momentum ($1<p_T<1.5$~GeV) and low mass from the
PHSD approach (red solid line) and the data of the PHENIX
Collaboration~\protect{\cite{Adare:2008ab}} shown by the black
symbols. The channel decomposition of the theoretical dilepton
spectrum is shown by the various colored lines (see legend).}
\label{dil}
\end{figure}

\begin{figure}
\resizebox{0.48\textwidth}{!}{%
 \includegraphics{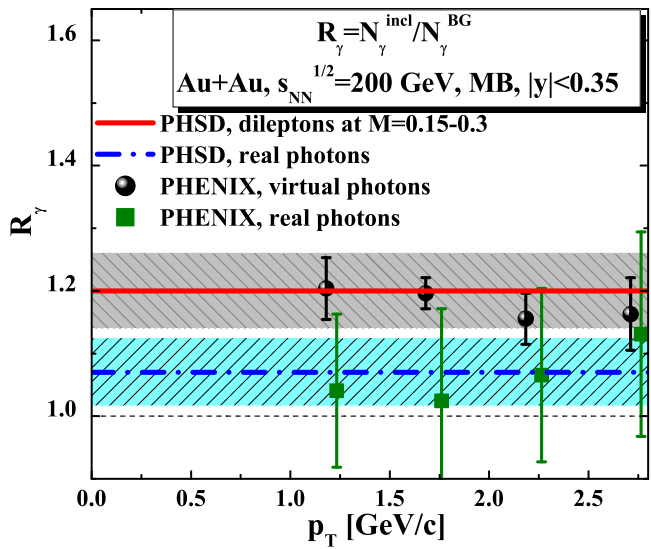}
} \caption{(Color online) The ratio of inclusive to background
photons as a function of $p_T$ from the PHSD in comparison to the
data of the PHENIX Collaboration~\protect{\cite{PHENIX1}}, using the
dilepton spectrum (red solid line) and the real photon yield
(dash-dotted blue line).} \label{Rg}
\end{figure}

In order to calculate the direct photon $v_2$ we need the calculated
inclusive photons $v_2$ (see the previous subsection), the flow of
the hadron decay background (we approximate
$v_2(\gamma^{BG})=v_2(\gamma^{\pi})$) and the quantity $R_\gamma$.
Let us remind that the direct photon spectrum was experimentally
obtained in Ref.~\cite{Adare:2008ab} by analyzing the yield of
dileptons with high transverse momentum $p_T$ and low invariant mass
$M$. We have studied the dilepton production at the top RHIC energy
within the PHSD approach in Ref.~\cite{Linnyk:2011vx}. The PHSD
results reproduced the PHENIX and STAR dilepton data differentially
in the invariant mass $M$ and transverse momentum $p_T$, only
underestimating the excess observed by PHENIX at low $M$ and low
$p_T$. Note, however, that for the relatively high transverse
momenta of dileptons $p_T>1$~GeV the agreement of the PHSD
calculations with the PHENIX data is quite good. We explicitly show
in Fig.~\ref{dil} the $p_T= \ $[1~GeV, 1.5~GeV] bin of the dilepton
spectrum from the PHSD versus the PHENIX data. This comparison is of
importance for the present investigation, since these data have been
used for the extraction of the ratio $R_\gamma$ by analyzing the
yield of dileptons in the invariant mass window $M=0.15-0.3$~GeV. We
present the resulting $R_\gamma$ from PHSD by the red solid line in
Fig.~\ref{Rg}.

Alternatively, we can use the calculated inclusive {\em real} photon
spectrum to find the ratio
$R_\gamma=N(\gamma)^{incl}/N(\gamma)^{BG}$. We present the
$R_\gamma$ as extracted from the real photons in PHSD  by the blue
dash-dotted line in Fig.~\ref{Rg}. The difference between the values
of $R_\gamma$ extracted from the dilepton spectra and the real
photon spectra is caused by the fact that in the dilepton mass
window $M=0.15-0.3$~GeV the background from the pion decays
effectively ``dies out", while the pion decay contribution is
prominent for $M\to0$, i.e. in the real photon spectrum.

We have followed the procedure of equation (\ref{dir1}) in the PHSD.
We obtain the red solid line in Fig.~\ref{v2_dir}, if we use the
$R_\gamma$ from the virtual photons in the invariant mass window
$M=0.15-0.3$~GeV, and the blue dash-dotted line, if we use the
$R_\gamma$ from the calculated real photon spectrum. The two lines
differ by about a factor of two.

Let us now consider an error analysis of the PHSD results. We keep in
mind that the statistical error in the calculated inclusive photon
$v_2$ is approximately $\delta v_2 (incl)=$15\%, cf.
Fig.~\ref{v2_incl}, and that of the background $v_2$ (which is a
combination of flow for the photons produced in hadronic decays, cf.
Fig.~\ref{v2_channels}) is about $\delta v_2 (BG)=$25\%. The error
of the ratio $R_\gamma$ is related to the accuracy of the dilepton
spectrum as in Fig.~\ref{dil}, which is approximately $\delta
R_\gamma=$25\%. From Eq.~(\ref{dir1}) we get for the error of $v_2
 (\gamma^{dir}) $ for fixed $p_T$,
\bea && \! \! \! \! \! \! \delta v_2(dir) =  \nn && \ \ \ \  \delta
v_2 (\gamma^{BG}) + \frac{R_\gamma}{ R_\gamma - 1} (\delta
v_2(\gamma^{BG}) + \delta v_2(\gamma^{incl})) \nn && \ \ \ \
 +
\delta R_\gamma \frac{v_2(\gamma^{incl}) - v_2
(\gamma^{BG})}{R_\gamma -1} \nn & & \ \ \ \
 +  \delta R_\gamma \frac{R_\gamma
}{(R_\gamma -1)^2} ( v_2(\gamma^{incl})-v_2(\gamma^{BG})), \eea
which we show as the upper (blue) error band in Fig.~\ref{v2_dir},
where the determination of $R_\gamma$ from the dilepton spectrum was
considered.

Both prescriptions for the direct photon $v_2$ have considerable
uncertainties (here we discussed only the statistical ones), and the
results agree within the errors with each other and with the data.
The PHENIX data on the direct photon $v_2$ can be described by the
PHSD calculations due to detailed modelling of the signal and the
background in photon as well as dilepton observables and by
following the same background subtraction procedure (driven by the
dilepton data in the mass window $M=0.15-0.3$~GeV) as in experiment.

We find in the direct photon $v_2$ no statistically significant evidence for
`new physics', beyond the radiations from the QGP and hadronic
matter produced in the collisions. On the other hand, the measured
large $v_2$ of inclusive and direct photons is a clear -- though
indirect -- signal of the QGP production in the early stages of the
collision. The strong interaction in the partonic medium is a
necessary prerequisite for the effective transfer of the collision
eccentricity into the asymmetry of the hadron momentum distribution
in the late stages, which in turn is reflected in the $v_2$ of the
produced photons. Indeed, the inclusive as well as direct photon
$v_2$ is underestimated in the purely hadronic scenario HSD.

\section{Conclusions}\label{sectionsummary}

In this study we have calculated the momentum spectra and the
elliptic flow $v_2$ of photons produced in minimal bias Au+Au
collisions at $\sqrt{s_{NN}}=200$~GeV using the microscopic PHSD
transport approach. For photon production we have incorporated the
interactions of quarks and gluons in the strongly interacting
quark-gluon plasma (sQGP) ($q+\bar q\to g+\gamma$ and
 $q(\bar q)+g\to q(\bar q)+\gamma$), the photon production in the hadronic decays
($\pi\to\gamma+\gamma$, $\eta\to\gamma+\gamma$,
$\omega\to\pi+\gamma$, $\eta'\to\rho+\gamma$, $\phi\to\eta+\gamma$,
$a_1\to\pi+\gamma$) as well as the hadronic interactions
($\pi+\pi\to\rho+\gamma$, $\rho+\pi\to\pi+\gamma$, and the
bremsstrahlung radiation $m+m\to m+m+\gamma$) of mesons produced
throughout the evolution of the collision.
We have calculated the photon production in the elementary off-shell
quark and gluon interactions by evaluating the tree-level diagrams
for the photon production in the scattering of massive, broad quarks
and gluons. The mesonic channels are treated using the same cross
sections as in the earlier HSD analysis of photon production at
lower collision energies~\cite{ElenaKiselev}.

We find that the PHSD calculations reproduce the transverse momentum
spectrum of direct photons as measured by the PHENIX Collaboration
in Refs.~\cite{PHENIXlast,Adare:2008ab}. Our microscopic
calculations access the channel decomposition of the observed direct
photon spectrum and show that the photons produced in the QGP
constitute slightly less than 50\% with the rest being distributed
among the other channels: mesonic interactions, decays of massive
hadronic resonances and the initial hard scatterings. Let us stress
that the dynamical calculations within the PHSD have reproduced the
measured differential spectra of dileptons produced in heavy-ion
collisions at SPS and RHIC energies (see
Refs.~\cite{Linnyk:2011hz,Linnyk:2011vx}). Also, it has been checked
in Ref.~\cite{Linnyk:2012pu} that the dilepton production from the
QGP constituents -- as incorporated in the
PHSD~\cite{olena2010,Linnyk:2011hz} -- agrees with the dilepton rate
emitted by the thermalized QCD medium as calculated in the lQCD
approach. We note, additionally, that the electric conductivity of
the QGP from the PHSD, which controls the photon emission rate in
equilibrium, is rather well in line with available lQCD results
\cite{Cassing:2013iz}.

We, furthermore, have demonstrated that the elliptic flow of pions
and the inclusive photon $v_2$ from PHSD are in a reasonable
agreement with the PHENIX data for the same observables. When
applying the same background substraction procedure for the 'direct
photon elliptic flow' as the PHENIX Collaboration (10) we find  a
rather good description of the direct photon $v_2$ (within errors).
However, an evaluation of the direct photon $v_2$ according to
Eq.(12) gives a lower signal, roughly in accord with the
computations in Refs.[3-7] although with sizeable error bars. The
difference between the two extraction procedures for the direct
photon flow $v_2$ can be attributed to different definitions for the
yield ratio of the inclusive and background photons ($R_\gamma$).

Our calculations show that the photon production in the QGP is
dominated by the early phase (similar to hydrodynamic models) and is
localized in the center of the fireball, where the collective flow
is still rather low, i.e. on the 2-3 \% level, only.
Thus, the strong $v_2$ of direct photons - which is comparable to
the hadronic $v_2$ - in PHSD is attributed to hadronic channels,
i.e. to meson binary reactions which are not subtracted in the
data. On the other hand, the strong $v_2$ of the 'parent' hadrons,
in turn, stems from the interactions in the QGP via collisions and
the partonic mean-filed potentials. Accordingly, the presence of the
QGP shows up 'indirectly' in the direct photon elliptic flow.

Finally, the high 'effective temperature' of the direct photons provides
strong evidence for a sizeable contribution of the photon emission from the QGP. The
experimental value of $T_{eff}(\mbox{exp})=233\pm19$~MeV can not be
explained by the photons of hadronic origin, even though the ``blue
shift" due to radial collective motion leads to a slope parameter
$T_{eff}(\mbox{hadrons})=200\pm20$~MeV, which is above the critical
temperature in our model $T_C=158$~MeV. On the other hand, the
partonic contribution to the photon spectrum has a considerably
higher $T_{eff}(\mbox{QGP})=260\pm20$~MeV. Taking into account both
the partonic and hadronic sources of photons, we obtain within PHSD
$T_{eff}(\mbox{PHSD})=220\pm20$~MeV and, therefore, reproduce the
measured 'effective temperature'. Our findings imply that there is
presently no clear signal for 'unconventional physics' (beyond the
strong interaction on the partonic and hadronic level) in the photon
data from the PHENIX Collaboration within error bars.


\section*{Acknowledgements}

The authors are grateful for fruitful discussions with G.~David,
A.~Drees, C.~Gale, H.~van Hees, C.M.~Ko, V.~Koch, L.~McLerran,
R.~Petti, R.~Rapp, V.~Skokov, E.~Shuryak, R.~Venugopalan and N.~Xu.
This study was supported by the LOEWE center HIC for FAIR and the
Margarete-Bieber Program at the University of Giessen.



\begin{thebibliography}{60}
\expandafter\ifx\csname natexlab\endcsname\relax\def\natexlab#1{#1}\fi
\expandafter\ifx\csname bibnamefont\endcsname\relax
  \def\bibnamefont#1{#1}\fi
\expandafter\ifx\csname bibfnamefont\endcsname\relax
  \def\bibfnamefont#1{#1}\fi
\expandafter\ifx\csname citenamefont\endcsname\relax
  \def\citenamefont#1{#1}\fi
\expandafter\ifx\csname url\endcsname\relax
  \def\url#1{\texttt{#1}}\fi
\expandafter\ifx\csname urlprefix\endcsname\relax\def\urlprefix{URL }\fi
\providecommand{\bibinfo}[2]{#2}
\providecommand{\eprint}[2][]{\url{#2}}

\bibitem[{\citenamefont{Peitzmann and Thoma}(2002)}]{Peitzmann:2001mz}
\bibinfo{author}{\bibfnamefont{T.}~\bibnamefont{Peitzmann}} \bibnamefont{and}
  \bibinfo{author}{\bibfnamefont{M.~H.} \bibnamefont{Thoma}},
  \bibinfo{journal}{Phys.Rept.} \textbf{\bibinfo{volume}{364}},
  \bibinfo{pages}{175} (\bibinfo{year}{2002}).

\bibitem[{\citenamefont{Adare et~al.}(2012)}]{PHENIX1}
\bibinfo{author}{\bibfnamefont{A.}~\bibnamefont{Adare}} \bibnamefont{et~al.}
  (\bibinfo{collaboration}{PHENIX Collaboration}),
  \bibinfo{journal}{Phys.Rev.Lett.} \textbf{\bibinfo{volume}{109}},
  \bibinfo{pages}{122302} (\bibinfo{year}{2012}).

\bibitem[{\citenamefont{Chatterjee et~al.}(2006)\citenamefont{Chatterjee,
  Frodermann, Heinz, and Srivastava}}]{Chatterjee:2005de}
\bibinfo{author}{\bibfnamefont{R.}~\bibnamefont{Chatterjee}},
  \bibinfo{author}{\bibfnamefont{E.~S.} \bibnamefont{Frodermann}},
  \bibinfo{author}{\bibfnamefont{U.~W.} \bibnamefont{Heinz}}, \bibnamefont{and}
  \bibinfo{author}{\bibfnamefont{D.~K.} \bibnamefont{Srivastava}},
  \bibinfo{journal}{Phys.Rev.Lett.} \textbf{\bibinfo{volume}{96}},
  \bibinfo{pages}{202302} (\bibinfo{year}{2006}).

\bibitem[{\citenamefont{Liu et~al.}(2009{\natexlab{a}})\citenamefont{Liu,
  Hirano, Werner, and Zhu}}]{Liu:2009kq}
\bibinfo{author}{\bibfnamefont{F.-M.} \bibnamefont{Liu}},
  \bibinfo{author}{\bibfnamefont{T.}~\bibnamefont{Hirano}},
  \bibinfo{author}{\bibfnamefont{K.}~\bibnamefont{Werner}}, \bibnamefont{and}
  \bibinfo{author}{\bibfnamefont{Y.}~\bibnamefont{Zhu}},
  \bibinfo{journal}{Nucl.Phys.} \textbf{\bibinfo{volume}{A830}},
  \bibinfo{pages}{587C} (\bibinfo{year}{2009}{\natexlab{a}}).

\bibitem[{\citenamefont{Dion et~al.}(2011{\natexlab{a}})\citenamefont{Dion,
  Gale, Jeon, Paquet, Schenke et~al.}}]{Dion:2011vd}
\bibinfo{author}{\bibfnamefont{M.}~\bibnamefont{Dion}},
  \bibinfo{author}{\bibfnamefont{C.}~\bibnamefont{Gale}},
  \bibinfo{author}{\bibfnamefont{S.}~\bibnamefont{Jeon}},
  \bibinfo{author}{\bibfnamefont{J.-F.} \bibnamefont{Paquet}},
  \bibinfo{author}{\bibfnamefont{B.}~\bibnamefont{Schenke}},
  \bibnamefont{et~al.}, \bibinfo{journal}{J.Phys.}
  \textbf{\bibinfo{volume}{G38}}, \bibinfo{pages}{124138}
  (\bibinfo{year}{2011}{\natexlab{a}}).

\bibitem[{\citenamefont{Dion et~al.}(2011{\natexlab{b}})\citenamefont{Dion,
  Paquet, Schenke, Young, Jeon et~al.}}]{Dion:2011pp}
\bibinfo{author}{\bibfnamefont{M.}~\bibnamefont{Dion}},
  \bibinfo{author}{\bibfnamefont{J.-F.} \bibnamefont{Paquet}},
  \bibinfo{author}{\bibfnamefont{B.}~\bibnamefont{Schenke}},
  \bibinfo{author}{\bibfnamefont{C.}~\bibnamefont{Young}},
  \bibinfo{author}{\bibfnamefont{S.}~\bibnamefont{Jeon}}, \bibnamefont{et~al.},
  \bibinfo{journal}{Phys.Rev.} \textbf{\bibinfo{volume}{C84}},
  \bibinfo{pages}{064901} (\bibinfo{year}{2011}{\natexlab{b}}).

\bibitem[{\citenamefont{Chatterjee et~al.}(2013)\citenamefont{Chatterjee,
  Holopainen, Helenius, Renk, and Eskola}}]{Chatterjee:2013naa}
\bibinfo{author}{\bibfnamefont{R.}~\bibnamefont{Chatterjee}},
  \bibinfo{author}{\bibfnamefont{H.}~\bibnamefont{Holopainen}},
  \bibinfo{author}{\bibfnamefont{I.}~\bibnamefont{Helenius}},
  \bibinfo{author}{\bibfnamefont{T.}~\bibnamefont{Renk}}, \bibnamefont{and}
  \bibinfo{author}{\bibfnamefont{K.~J.} \bibnamefont{Eskola}}
  (\bibinfo{year}{2013}), \eprint{arXiv: 1305.6443}.

\bibitem[{\citenamefont{Bzdak and Skokov}(2013)}]{Bzdak:2012fr}
\bibinfo{author}{\bibfnamefont{A.}~\bibnamefont{Bzdak}} \bibnamefont{and}
  \bibinfo{author}{\bibfnamefont{V.}~\bibnamefont{Skokov}},
  \bibinfo{journal}{Phys.Rev.Lett.} \textbf{\bibinfo{volume}{110}},
  \bibinfo{pages}{192301} (\bibinfo{year}{2013}).

\bibitem[{\citenamefont{Basar et~al.}(2012)\citenamefont{Basar, Kharzeev,
  Kharzeev, and Skokov}}]{Basar:2012bp}
\bibinfo{author}{\bibfnamefont{G.}~\bibnamefont{Basar}},
  \bibinfo{author}{\bibfnamefont{D.}~\bibnamefont{Kharzeev}},
  \bibinfo{author}{\bibfnamefont{D.}~\bibnamefont{Kharzeev}}, \bibnamefont{and}
  \bibinfo{author}{\bibfnamefont{V.}~\bibnamefont{Skokov}},
  \bibinfo{journal}{Phys.Rev.Lett.} \textbf{\bibinfo{volume}{109}},
  \bibinfo{pages}{202303} (\bibinfo{year}{2012}).

\bibitem[{\citenamefont{Goloviznin et~al.}(2012)\citenamefont{Goloviznin,
  Snigirev, and Zinovjev}}]{Goloviznin:2012dy}
\bibinfo{author}{\bibfnamefont{V.}~\bibnamefont{Goloviznin}},
  \bibinfo{author}{\bibfnamefont{A.}~\bibnamefont{Snigirev}}, \bibnamefont{and}
  \bibinfo{author}{\bibfnamefont{G.}~\bibnamefont{Zinovjev}}
  (\bibinfo{year}{2012}), \eprint{arXiv:1209.2380}.

\bibitem[{\citenamefont{Pantuev}(2011)}]{Pantuev:2011yh}
\bibinfo{author}{\bibfnamefont{V.}~\bibnamefont{Pantuev}}
  (\bibinfo{year}{2011}), \eprint{arXiv: 1105.4033}.

\bibitem[{\citenamefont{van Hees et~al.}(2011)\citenamefont{van Hees, Gale, and
  Rapp}}]{vanHees:2011vb}
\bibinfo{author}{\bibfnamefont{H.}~\bibnamefont{van Hees}},
  \bibinfo{author}{\bibfnamefont{C.}~\bibnamefont{Gale}}, \bibnamefont{and}
  \bibinfo{author}{\bibfnamefont{R.}~\bibnamefont{Rapp}},
  \bibinfo{journal}{Phys.Rev.} \textbf{\bibinfo{volume}{C84}},
  \bibinfo{pages}{054906} (\bibinfo{year}{2011}).

\bibitem[{\citenamefont{Bratkovskaya
  et~al.}(2011{\natexlab{a}})\citenamefont{Bratkovskaya, Cassing, Konchakovski,
  and Linnyk}}]{PHSDqscaling}
\bibinfo{author}{\bibfnamefont{E.}~\bibnamefont{Bratkovskaya}},
  \bibinfo{author}{\bibfnamefont{W.}~\bibnamefont{Cassing}},
  \bibinfo{author}{\bibfnamefont{V.}~\bibnamefont{Konchakovski}},
  \bibnamefont{and} \bibinfo{author}{\bibfnamefont{O.}~\bibnamefont{Linnyk}},
  \bibinfo{journal}{Nucl.Phys.} \textbf{\bibinfo{volume}{A856}},
  \bibinfo{pages}{162} (\bibinfo{year}{2011}{\natexlab{a}}).

\bibitem[{\citenamefont{Linnyk et~al.}(2011{\natexlab{a}})\citenamefont{Linnyk,
  Bratkovskaya, Ozvenchuk, Cassing, and Ko}}]{Linnyk:2011hz}
\bibinfo{author}{\bibfnamefont{O.}~\bibnamefont{Linnyk}},
  \bibinfo{author}{\bibfnamefont{E.~L.} \bibnamefont{Bratkovskaya}},
  \bibinfo{author}{\bibfnamefont{V.}~\bibnamefont{Ozvenchuk}},
  \bibinfo{author}{\bibfnamefont{W.}~\bibnamefont{Cassing}}, \bibnamefont{and}
  \bibinfo{author}{\bibfnamefont{C.~M.} \bibnamefont{Ko}},
  \bibinfo{journal}{Phys.Rev.} \textbf{\bibinfo{volume}{C84}},
  \bibinfo{pages}{054917} (\bibinfo{year}{2011}{\natexlab{a}}).

\bibitem[{\citenamefont{Linnyk et~al.}(2012)\citenamefont{Linnyk, Cassing,
  Manninen, Bratkovskaya, and Ko}}]{Linnyk:2011vx}
\bibinfo{author}{\bibfnamefont{O.}~\bibnamefont{Linnyk}},
  \bibinfo{author}{\bibfnamefont{W.}~\bibnamefont{Cassing}},
  \bibinfo{author}{\bibfnamefont{J.}~\bibnamefont{Manninen}},
  \bibinfo{author}{\bibfnamefont{E.}~\bibnamefont{Bratkovskaya}},
  \bibnamefont{and} \bibinfo{author}{\bibfnamefont{C.}~\bibnamefont{Ko}},
  \bibinfo{journal}{Phys.Rev.} \textbf{\bibinfo{volume}{C85}},
  \bibinfo{pages}{024910} (\bibinfo{year}{2012}).

\bibitem[{\citenamefont{Linnyk et~al.}(2013)\citenamefont{Linnyk, Cassing,
  Manninen, Bratkovskaya, Gossiaux et~al.}}]{Linnyk:2012pu}
\bibinfo{author}{\bibfnamefont{O.}~\bibnamefont{Linnyk}},
  \bibinfo{author}{\bibfnamefont{W.}~\bibnamefont{Cassing}},
  \bibinfo{author}{\bibfnamefont{J.}~\bibnamefont{Manninen}},
  \bibinfo{author}{\bibfnamefont{E.}~\bibnamefont{Bratkovskaya}},
  \bibinfo{author}{\bibfnamefont{P.}~\bibnamefont{Gossiaux}},
  \bibnamefont{et~al.}, \bibinfo{journal}{Phys.Rev.}
  \textbf{\bibinfo{volume}{C87}}, \bibinfo{pages}{014905}
  (\bibinfo{year}{2013}).

\bibitem[{\citenamefont{Cassing and Bratkovskaya}(1999)}]{Cass99}
\bibinfo{author}{\bibfnamefont{W.}~\bibnamefont{Cassing}} \bibnamefont{and}
  \bibinfo{author}{\bibfnamefont{E.~L.} \bibnamefont{Bratkovskaya}},
  \bibinfo{journal}{Phys. Rept.} \textbf{\bibinfo{volume}{308}},
  \bibinfo{pages}{65} (\bibinfo{year}{1999}).

\bibitem[{\citenamefont{Bratkovskaya and Cassing}(1997)}]{Brat97}
\bibinfo{author}{\bibfnamefont{E.~L.} \bibnamefont{Bratkovskaya}}
  \bibnamefont{and} \bibinfo{author}{\bibfnamefont{W.}~\bibnamefont{Cassing}},
  \bibinfo{journal}{Nucl. Phys.} \textbf{\bibinfo{volume}{A 619}},
  \bibinfo{pages}{413} (\bibinfo{year}{1997}).

\bibitem[{\citenamefont{Ehehalt and Cassing}(1996)}]{Ehehalt}
\bibinfo{author}{\bibfnamefont{W.}~\bibnamefont{Ehehalt}} \bibnamefont{and}
  \bibinfo{author}{\bibfnamefont{W.}~\bibnamefont{Cassing}},
  \bibinfo{journal}{Nucl. Phys.} \textbf{\bibinfo{volume}{A 602}},
  \bibinfo{pages}{449} (\bibinfo{year}{1996}).

\bibitem[{\citenamefont{Bratkovskaya et~al.}(2008)\citenamefont{Bratkovskaya,
  Kiselev, and Sharkov}}]{ElenaKiselev}
\bibinfo{author}{\bibfnamefont{E.}~\bibnamefont{Bratkovskaya}},
  \bibinfo{author}{\bibfnamefont{S.}~\bibnamefont{Kiselev}}, \bibnamefont{and}
  \bibinfo{author}{\bibfnamefont{G.}~\bibnamefont{Sharkov}},
  \bibinfo{journal}{Phys.Rev.} \textbf{\bibinfo{volume}{C78}},
  \bibinfo{pages}{034905} (\bibinfo{year}{2008}).

\bibitem[{\citenamefont{Adams et~al.}(2005{\natexlab{a}})}]{STARQGP}
\bibinfo{author}{\bibfnamefont{J.}~\bibnamefont{Adams}} \bibnamefont{et~al.}
  (\bibinfo{collaboration}{STAR Collaboration}), \bibinfo{journal}{Nucl.Phys.}
  \textbf{\bibinfo{volume}{A757}}, \bibinfo{pages}{102}
  (\bibinfo{year}{2005}{\natexlab{a}}).

\bibitem[{\citenamefont{Adcox et~al.}(2005)}]{PHENIXQGP}
\bibinfo{author}{\bibfnamefont{K.}~\bibnamefont{Adcox}} \bibnamefont{et~al.}
  (\bibinfo{collaboration}{PHENIX Collaboration}),
  \bibinfo{journal}{Nucl.Phys.} \textbf{\bibinfo{volume}{A757}},
  \bibinfo{pages}{184} (\bibinfo{year}{2005}).

\bibitem[{\citenamefont{Arsene et~al.}(2005)}]{BRAHMS}
\bibinfo{author}{\bibfnamefont{I.}~\bibnamefont{Arsene}} \bibnamefont{et~al.}
  (\bibinfo{collaboration}{BRAHMS Collaboration}),
  \bibinfo{journal}{Nucl.Phys.} \textbf{\bibinfo{volume}{A757}},
  \bibinfo{pages}{1} (\bibinfo{year}{2005}).

\bibitem[{\citenamefont{Back et~al.}(2005)\citenamefont{Back, Baker,
  Ballintijn, Barton, Becker et~al.}}]{PHOBOS}
\bibinfo{author}{\bibfnamefont{B.}~\bibnamefont{Back}},
  \bibinfo{author}{\bibfnamefont{M.}~\bibnamefont{Baker}},
  \bibinfo{author}{\bibfnamefont{M.}~\bibnamefont{Ballintijn}},
  \bibinfo{author}{\bibfnamefont{D.}~\bibnamefont{Barton}},
  \bibinfo{author}{\bibfnamefont{B.}~\bibnamefont{Becker}},
  \bibnamefont{et~al.}, \bibinfo{journal}{Nucl.Phys.}
  \textbf{\bibinfo{volume}{A757}}, \bibinfo{pages}{28} (\bibinfo{year}{2005}).

\bibitem[{\citenamefont{Bratkovskaya
  et~al.}(2011{\natexlab{b}})\citenamefont{Bratkovskaya, Cassing, Konchakovski,
  and Linnyk}}]{BrCa11}
\bibinfo{author}{\bibfnamefont{E.~L.} \bibnamefont{Bratkovskaya}},
  \bibinfo{author}{\bibfnamefont{W.}~\bibnamefont{Cassing}},
  \bibinfo{author}{\bibfnamefont{V.~P.} \bibnamefont{Konchakovski}},
  \bibnamefont{and} \bibinfo{author}{\bibfnamefont{O.}~\bibnamefont{Linnyk}},
  \bibinfo{journal}{Nucl. Phys.} \textbf{\bibinfo{volume}{A856}},
  \bibinfo{pages}{162} (\bibinfo{year}{2011}{\natexlab{b}}).

\bibitem[{\citenamefont{Feinberg}(1976)}]{Feinberg:1976ua}
\bibinfo{author}{\bibfnamefont{E.~L.} \bibnamefont{Feinberg}},
  \bibinfo{journal}{Nuovo Cim.} \textbf{\bibinfo{volume}{A34}},
  \bibinfo{pages}{391} (\bibinfo{year}{1976}).

\bibitem[{\citenamefont{Shuryak}(1978)}]{Shuryak:1978ij}
\bibinfo{author}{\bibfnamefont{E.~V.} \bibnamefont{Shuryak}},
  \bibinfo{journal}{Phys. Lett.} \textbf{\bibinfo{volume}{B78}},
  \bibinfo{pages}{150} (\bibinfo{year}{1978}),
  \bibinfo{note}{{Sov.~J.~Nucl.~Phys. {\bf 28} (1978) 408, Yad.~Fiz. {\bf 28}
  (1978) 796}}.

\bibitem[{\citenamefont{Kapusta et~al.}(1991)\citenamefont{Kapusta, Lichard,
  and Seibert}}]{Kapusta:1991qp}
\bibinfo{author}{\bibfnamefont{J.~I.} \bibnamefont{Kapusta}},
  \bibinfo{author}{\bibfnamefont{P.}~\bibnamefont{Lichard}}, \bibnamefont{and}
  \bibinfo{author}{\bibfnamefont{D.}~\bibnamefont{Seibert}},
  \bibinfo{journal}{Phys.Rev.} \textbf{\bibinfo{volume}{D44}},
  \bibinfo{pages}{2774} (\bibinfo{year}{1991}).

\bibitem[{\citenamefont{Alam et~al.}(2001)\citenamefont{Alam, Sarkar, Hatsuda,
  Nayak, and Sinha}}]{Alam:2000bu}
\bibinfo{author}{\bibfnamefont{J.-e.} \bibnamefont{Alam}},
  \bibinfo{author}{\bibfnamefont{S.}~\bibnamefont{Sarkar}},
  \bibinfo{author}{\bibfnamefont{T.}~\bibnamefont{Hatsuda}},
  \bibinfo{author}{\bibfnamefont{T.~K.} \bibnamefont{Nayak}}, \bibnamefont{and}
  \bibinfo{author}{\bibfnamefont{B.}~\bibnamefont{Sinha}},
  \bibinfo{journal}{Phys.Rev.} \textbf{\bibinfo{volume}{C63}},
  \bibinfo{pages}{021901} (\bibinfo{year}{2001}).

\bibitem[{\citenamefont{Steffen and Thoma}(2001)}]{Steffen:2001pv}
\bibinfo{author}{\bibfnamefont{F.~D.} \bibnamefont{Steffen}} \bibnamefont{and}
  \bibinfo{author}{\bibfnamefont{M.~H.} \bibnamefont{Thoma}},
  \bibinfo{journal}{Phys.Lett.} \textbf{\bibinfo{volume}{B510}},
  \bibinfo{pages}{98} (\bibinfo{year}{2001}).

\bibitem[{\citenamefont{Srivastava and Sinha}(2001)}]{Srivastava:2000pv}
\bibinfo{author}{\bibfnamefont{D.~K.} \bibnamefont{Srivastava}}
  \bibnamefont{and} \bibinfo{author}{\bibfnamefont{B.}~\bibnamefont{Sinha}},
  \bibinfo{journal}{Phys.Rev.} \textbf{\bibinfo{volume}{C64}},
  \bibinfo{pages}{034902} (\bibinfo{year}{2001}).

\bibitem[{\citenamefont{Huovinen et~al.}(2002)\citenamefont{Huovinen,
  Ruuskanen, and Rasanen}}]{Huovinen:2001wx}
\bibinfo{author}{\bibfnamefont{P.}~\bibnamefont{Huovinen}},
  \bibinfo{author}{\bibfnamefont{P.}~\bibnamefont{Ruuskanen}},
  \bibnamefont{and} \bibinfo{author}{\bibfnamefont{S.}~\bibnamefont{Rasanen}},
  \bibinfo{journal}{Phys.Lett.} \textbf{\bibinfo{volume}{B535}},
  \bibinfo{pages}{109} (\bibinfo{year}{2002}).

\bibitem[{\citenamefont{Turbide et~al.}(2004)\citenamefont{Turbide, Rapp, and
  Gale}}]{Turbide:2003si}
\bibinfo{author}{\bibfnamefont{S.}~\bibnamefont{Turbide}},
  \bibinfo{author}{\bibfnamefont{R.}~\bibnamefont{Rapp}}, \bibnamefont{and}
  \bibinfo{author}{\bibfnamefont{C.}~\bibnamefont{Gale}},
  \bibinfo{journal}{Phys.Rev.} \textbf{\bibinfo{volume}{C69}},
  \bibinfo{pages}{014903} (\bibinfo{year}{2004}).

\bibitem[{\citenamefont{d'Enterria and Peressounko}(2006)}]{d'Enterria:2005vz}
\bibinfo{author}{\bibfnamefont{D.~G.} \bibnamefont{d'Enterria}}
  \bibnamefont{and}
  \bibinfo{author}{\bibfnamefont{D.}~\bibnamefont{Peressounko}},
  \bibinfo{journal}{Eur.Phys.J.} \textbf{\bibinfo{volume}{C46}},
  \bibinfo{pages}{451} (\bibinfo{year}{2006}).

\bibitem[{\citenamefont{Liu et~al.}(2009{\natexlab{b}})\citenamefont{Liu,
  Hirano, Werner, and Zhu}}]{Liu:2008eh}
\bibinfo{author}{\bibfnamefont{F.-M.} \bibnamefont{Liu}},
  \bibinfo{author}{\bibfnamefont{T.}~\bibnamefont{Hirano}},
  \bibinfo{author}{\bibfnamefont{K.}~\bibnamefont{Werner}}, \bibnamefont{and}
  \bibinfo{author}{\bibfnamefont{Y.}~\bibnamefont{Zhu}},
  \bibinfo{journal}{Phys.Rev.} \textbf{\bibinfo{volume}{C79}},
  \bibinfo{pages}{014905} (\bibinfo{year}{2009}{\natexlab{b}}).

\bibitem[{\citenamefont{Turbide et~al.}(2008)\citenamefont{Turbide, Gale,
  Frodermann, and Heinz}}]{Turbide:2007mi}
\bibinfo{author}{\bibfnamefont{S.}~\bibnamefont{Turbide}},
  \bibinfo{author}{\bibfnamefont{C.}~\bibnamefont{Gale}},
  \bibinfo{author}{\bibfnamefont{E.}~\bibnamefont{Frodermann}},
  \bibnamefont{and} \bibinfo{author}{\bibfnamefont{U.}~\bibnamefont{Heinz}},
  \bibinfo{journal}{Phys.Rev.} \textbf{\bibinfo{volume}{C77}},
  \bibinfo{pages}{024909} (\bibinfo{year}{2008}).

\bibitem[{\citenamefont{Adare et~al.}(2010{\natexlab{a}})}]{PHENIXlast}
\bibinfo{author}{\bibfnamefont{A.}~\bibnamefont{Adare}} \bibnamefont{et~al.}
  (\bibinfo{collaboration}{PHENIX}), \bibinfo{journal}{Phys. Rev.}
  \textbf{\bibinfo{volume}{C 81}}, \bibinfo{pages}{034911}
  (\bibinfo{year}{2010}{\natexlab{a}}).

\bibitem[{\citenamefont{Adare et~al.}(2010{\natexlab{b}})}]{Adare:2008ab}
\bibinfo{author}{\bibfnamefont{A.}~\bibnamefont{Adare}} \bibnamefont{et~al.}
  (\bibinfo{collaboration}{PHENIX Collaboration}),
  \bibinfo{journal}{Phys.Rev.Lett.} \textbf{\bibinfo{volume}{104}},
  \bibinfo{pages}{132301} (\bibinfo{year}{2010}{\natexlab{b}}).

\bibitem[{\citenamefont{Bratkovskaya et~al.}(2009)\citenamefont{Bratkovskaya,
  Cassing, and Linnyk}}]{Bratkovskaya:2008bf}
\bibinfo{author}{\bibfnamefont{E.~L.} \bibnamefont{Bratkovskaya}},
  \bibinfo{author}{\bibfnamefont{W.}~\bibnamefont{Cassing}}, \bibnamefont{and}
  \bibinfo{author}{\bibfnamefont{O.}~\bibnamefont{Linnyk}},
  \bibinfo{journal}{Phys. Lett.} \textbf{\bibinfo{volume}{B670}},
  \bibinfo{pages}{428} (\bibinfo{year}{2009}).

\bibitem[{\citenamefont{Cassing and Bratkovskaya}(2009)}]{CasBrat}
\bibinfo{author}{\bibfnamefont{W.}~\bibnamefont{Cassing}} \bibnamefont{and}
  \bibinfo{author}{\bibfnamefont{E.~L.} \bibnamefont{Bratkovskaya}},
  \bibinfo{journal}{Nucl. Phys.} \textbf{\bibinfo{volume}{A 831}},
  \bibinfo{pages}{215} (\bibinfo{year}{2009}).

\bibitem[{\citenamefont{Cassing and Bratkovskaya}(2008)}]{Cass08}
\bibinfo{author}{\bibfnamefont{W.}~\bibnamefont{Cassing}} \bibnamefont{and}
  \bibinfo{author}{\bibfnamefont{E.~L.} \bibnamefont{Bratkovskaya}},
  \bibinfo{journal}{Phys. Rev.} \textbf{\bibinfo{volume}{C 78}},
  \bibinfo{pages}{034919} (\bibinfo{year}{2008}).

\bibitem[{\citenamefont{Cassing and Juchem}(2000)}]{Cass_off1}
\bibinfo{author}{\bibfnamefont{W.}~\bibnamefont{Cassing}} \bibnamefont{and}
  \bibinfo{author}{\bibfnamefont{S.}~\bibnamefont{Juchem}},
  \bibinfo{journal}{Nucl. Phys.} \textbf{\bibinfo{volume}{A 665}},
  \bibinfo{pages}{377} (\bibinfo{year}{2000}), \bibinfo{note}{{\it ibid.} { A
  }{\bf 672}, 417 (2000)}.

\bibitem[{\citenamefont{Cassing}(2009)}]{Cassing:2008nn}
\bibinfo{author}{\bibfnamefont{W.}~\bibnamefont{Cassing}},
  \bibinfo{journal}{Eur. Phys. J. ST} \textbf{\bibinfo{volume}{168}},
  \bibinfo{pages}{3} (\bibinfo{year}{2009}).

\bibitem[{\citenamefont{Linnyk et~al.}(2011{\natexlab{b}})\citenamefont{Linnyk,
  Bratkovskaya, Manninen, and Cassing}}]{Linnyk:2011ee}
\bibinfo{author}{\bibfnamefont{O.}~\bibnamefont{Linnyk}},
  \bibinfo{author}{\bibfnamefont{E.}~\bibnamefont{Bratkovskaya}},
  \bibinfo{author}{\bibfnamefont{J.}~\bibnamefont{Manninen}}, \bibnamefont{and}
  \bibinfo{author}{\bibfnamefont{W.}~\bibnamefont{Cassing}},
  \bibinfo{journal}{J.Phys.Conf.Ser.} \textbf{\bibinfo{volume}{312}},
  \bibinfo{pages}{012010} (\bibinfo{year}{2011}{\natexlab{b}}).

\bibitem[{\citenamefont{Beringer et~al.}(2012)}]{PDG}
\bibinfo{author}{\bibfnamefont{J.}~\bibnamefont{Beringer}} \bibnamefont{et~al.}
  (\bibinfo{collaboration}{Particle Data Group}), \bibinfo{journal}{Phys.Rev.}
  \textbf{\bibinfo{volume}{D86}}, \bibinfo{pages}{010001}
  (\bibinfo{year}{2012}).

\bibitem[{\citenamefont{Gale and Kapusta}(1987)}]{Gale87}
\bibinfo{author}{\bibfnamefont{C.}~\bibnamefont{Gale}} \bibnamefont{and}
  \bibinfo{author}{\bibfnamefont{J.}~\bibnamefont{Kapusta}},
  \bibinfo{journal}{Phys. Rev. C} \textbf{\bibinfo{volume}{35}},
  \bibinfo{pages}{2107} (\bibinfo{year}{1987}), \bibinfo{note}{{\it idem},
  Phys. Rev. C {\bf 38}, 2659 (1988), Nucl. Phys. A {\bf 495}, 423c (1989)}.

\bibitem[{\citenamefont{Eggers et~al.}(1996)\citenamefont{Eggers, Tabti, Gale,
  and Haglin}}]{PhysRevD.53.4822}
\bibinfo{author}{\bibfnamefont{H.~C.} \bibnamefont{Eggers}},
  \bibinfo{author}{\bibfnamefont{R.}~\bibnamefont{Tabti}},
  \bibinfo{author}{\bibfnamefont{C.}~\bibnamefont{Gale}}, \bibnamefont{and}
  \bibinfo{author}{\bibfnamefont{K.}~\bibnamefont{Haglin}},
  \bibinfo{journal}{Phys. Rev. D} \textbf{\bibinfo{volume}{53}},
  \bibinfo{pages}{4822} (\bibinfo{year}{1996}).

\bibitem[{\citenamefont{Song et~al.}(1994)\citenamefont{Song, Ko, and
  Gale}}]{Song:1994zs}
\bibinfo{author}{\bibfnamefont{C.}~\bibnamefont{Song}},
  \bibinfo{author}{\bibfnamefont{C.~M.} \bibnamefont{Ko}}, \bibnamefont{and}
  \bibinfo{author}{\bibfnamefont{C.}~\bibnamefont{Gale}},
  \bibinfo{journal}{Phys. Rev.} \textbf{\bibinfo{volume}{D50}},
  \bibinfo{pages}{1827} (\bibinfo{year}{1994}).

\bibitem[{\citenamefont{Song}(1993)}]{Song:1993ae}
\bibinfo{author}{\bibfnamefont{C.}~\bibnamefont{Song}},
  \bibinfo{journal}{Phys.Rev.} \textbf{\bibinfo{volume}{C47}},
  \bibinfo{pages}{2861} (\bibinfo{year}{1993}).

\bibitem[{\citenamefont{Marzani and Ball}(2009)}]{Marzani:2008uh}
\bibinfo{author}{\bibfnamefont{S.}~\bibnamefont{Marzani}} \bibnamefont{and}
  \bibinfo{author}{\bibfnamefont{R.~D.} \bibnamefont{Ball}},
  \bibinfo{journal}{Nucl.Phys.} \textbf{\bibinfo{volume}{B814}},
  \bibinfo{pages}{246} (\bibinfo{year}{2009}).

\bibitem[{\citenamefont{Wong and Wang}(1998)}]{Wong:1998pq}
\bibinfo{author}{\bibfnamefont{C.-Y.} \bibnamefont{Wong}} \bibnamefont{and}
  \bibinfo{author}{\bibfnamefont{H.}~\bibnamefont{Wang}},
  \bibinfo{journal}{Phys.Rev.} \textbf{\bibinfo{volume}{C58}},
  \bibinfo{pages}{376} (\bibinfo{year}{1998}).

\bibitem[{\citenamefont{Linnyk et~al.}(2005)\citenamefont{Linnyk, Leupold, and
  Mosel}}]{Linnyk:2004mt}
\bibinfo{author}{\bibfnamefont{O.}~\bibnamefont{Linnyk}},
  \bibinfo{author}{\bibfnamefont{S.}~\bibnamefont{Leupold}}, \bibnamefont{and}
  \bibinfo{author}{\bibfnamefont{U.}~\bibnamefont{Mosel}},
  \bibinfo{journal}{Phys. Rev.} \textbf{\bibinfo{volume}{D71}},
  \bibinfo{pages}{034009} (\bibinfo{year}{2005}).

\bibitem[{\citenamefont{Linnyk}(2011)}]{olena2010}
\bibinfo{author}{\bibfnamefont{O.}~\bibnamefont{Linnyk}}, \bibinfo{journal}{J.
  Phys.} \textbf{\bibinfo{volume}{G38}}, \bibinfo{pages}{025105}
  (\bibinfo{year}{2011}).

\bibitem[{\citenamefont{Borsanyi et~al.}(2010)}]{Borsanyi:2010bp}
\bibinfo{author}{\bibfnamefont{S.}~\bibnamefont{Borsanyi}} \bibnamefont{et~al.}
  (\bibinfo{collaboration}{Wuppertal-Budapest Collaboration}),
  \bibinfo{journal}{JHEP} \textbf{\bibinfo{volume}{1009}}, \bibinfo{pages}{073}
  (\bibinfo{year}{2010}).

\bibitem[{\citenamefont{Adams et~al.}(2005{\natexlab{b}})}]{Adams:2004bi}
\bibinfo{author}{\bibfnamefont{J.}~\bibnamefont{Adams}} \bibnamefont{et~al.}
  (\bibinfo{collaboration}{STAR Collaboration}), \bibinfo{journal}{Phys.Rev.}
  \textbf{\bibinfo{volume}{C72}}, \bibinfo{pages}{014904}
  (\bibinfo{year}{2005}{\natexlab{b}}).

\bibitem[{\citenamefont{Adler et~al.}(2003)}]{Adler:2003kt}
\bibinfo{author}{\bibfnamefont{S.}~\bibnamefont{Adler}} \bibnamefont{et~al.}
  (\bibinfo{collaboration}{PHENIX Collaboration}),
  \bibinfo{journal}{Phys.Rev.Lett.} \textbf{\bibinfo{volume}{91}},
  \bibinfo{pages}{182301} (\bibinfo{year}{2003}).

\bibitem[{\citenamefont{Konchakovski et~al.}(2012)\citenamefont{Konchakovski,
  Bratkovskaya, Cassing, Toneev, Voloshin et~al.}}]{PHSDasymmetries}
\bibinfo{author}{\bibfnamefont{V.}~\bibnamefont{Konchakovski}},
  \bibinfo{author}{\bibfnamefont{E.}~\bibnamefont{Bratkovskaya}},
  \bibinfo{author}{\bibfnamefont{W.}~\bibnamefont{Cassing}},
  \bibinfo{author}{\bibfnamefont{V.}~\bibnamefont{Toneev}},
  \bibinfo{author}{\bibfnamefont{S.}~\bibnamefont{Voloshin}},
  \bibnamefont{et~al.}, \bibinfo{journal}{Phys.Rev.}
  \textbf{\bibinfo{volume}{C85}}, \bibinfo{pages}{044922}
  (\bibinfo{year}{2012}).

\bibitem[{\citenamefont{David}(2013)}]{Gaboprivat}
\bibinfo{author}{\bibfnamefont{G.}~\bibnamefont{David}} (\bibinfo{year}{2013}),
  \bibinfo{note}{private communication}.

\bibitem[{\citenamefont{Tserruya}(2013)}]{Tserruya:2012jb}
\bibinfo{author}{\bibfnamefont{I.}~\bibnamefont{Tserruya}}
  (\bibinfo{collaboration}{PHENIX Collaboration}),
  \bibinfo{journal}{Nucl.Phys.} \textbf{\bibinfo{volume}{A904-905}},
  \bibinfo{pages}{225c} (\bibinfo{year}{2013}).

\bibitem[{\citenamefont{Cassing et~al.}(2013)\citenamefont{Cassing, Linnyk,
  Steinert, and Ozvenchuk}}]{Cassing:2013iz}
\bibinfo{author}{\bibfnamefont{W.}~\bibnamefont{Cassing}},
  \bibinfo{author}{\bibfnamefont{O.}~\bibnamefont{Linnyk}},
  \bibinfo{author}{\bibfnamefont{T.}~\bibnamefont{Steinert}}, \bibnamefont{and}
  \bibinfo{author}{\bibfnamefont{V.}~\bibnamefont{Ozvenchuk}},
  \bibinfo{journal}{Phys.Rev.Lett.} \textbf{\bibinfo{volume}{110}},
  \bibinfo{pages}{182301} (\bibinfo{year}{2013}).

\end{thebibliography}

\end{document}